\documentclass[review,number]{elsarticle}
\biboptions{sort&compress}

\usepackage{lineno}
\usepackage[version=4]{mhchem} 
\usepackage[english]{babel}
\usepackage[utf8]{inputenc}
\usepackage[T1]{fontenc}
\usepackage{amsmath}
\usepackage{graphicx}
\usepackage{tabularx} 
\usepackage{threeparttable,booktabs}
\usepackage{array, collcell}
\usepackage{multirow}
\usepackage{url}
\usepackage{float}
\usepackage{siunitx}

\usepackage{xcolor}
\usepackage{listings}

\newcommand\AddLabel[1]{%
  \refstepcounter{equation}
  (\theequation)
  \label{#1}
}
\newcolumntype{M}{>{$\displaystyle}X<{$\hfil}} 
\newcolumntype{L}{>{\collectcell\AddLabel}r<{\endcollectcell}}

\DeclareSIUnit\angstrom{\text {Å}}
\DeclareSIUnit\kcal{kcal}
\DeclareSIUnit\year{year}
\DeclareSIUnit\GWh{GWh}
\DeclareUnicodeCharacter{2212}{-}
\begin{document}

\begin{frontmatter}

\journal{arXiv}

\title{Lithium-ion battery modelling for nonisothermal conditions}

\author[1,3]{Felix Schloms}    
\affiliation[1]{organization={PoreLab, Department of Chemistry, Norwegian University of Science and Technology, NTNU}, postcode={7491}, city={Trondheim}, country={Norway}}
\affiliation[3]{organization={Present address: Institute of Technical Thermodynamics and Thermal Process Engineering, University of Stuttgart}, postcode={70569}, city={Stuttgart}, country={Germany}}
\author[2]{Øystein Gullbrekken\corref{cor1}}
\affiliation[2]{organization={Department of Materials Science and Engineering, Norwegian University of Science and Technology, NTNU}, postcode={7491}, city={Trondheim}, country={Norway}}
\ead{oystein.gullbrekken@ntnu.no}
\author[1]{Signe Kjelstrup}

\cortext[cor1]{Corresponding author}

\begin{abstract}
A nonequilibrium thermodynamic model is presented for the nonisothermal lithium-ion battery cell. Coupling coefficients, all significant for transport of heat, mass, charge and chemical reaction, were used to model profiles of temperature, concentration and electric potential for each layer of the cell. Electrode surfaces were modelled with excess properties. Extending earlier works, we included lithium diffusion in the electrodes, and explained the cell's thermal signature due to Peltier and Soret effects.

We showed that the model is consistent with the second law of thermodynamics, meaning that the entropy production computed at steady state from entropy fluxes is equal to the integral over the sum of flux-force products. The procedure is beneficial in electrochemical cell modelling as it reveals inconsistencies.

The model was solved for typical lithium-ion battery materials. The coupling coefficients for transport of salts and solvents lead to significant concentration polarization. Thermal polarization is then negligible. We show that a zero-valued heat flux is not necessarily synonymous with a zero temperature gradient. Results are important for efforts that aim to avoid local hot spots. A program code is made available for testing and applications. The program is designed to solve dynamic boundary value problems posed by the electrode surfaces.


\end{abstract}




\end{frontmatter}

\section{Introduction}

Electric energy storage is essential for realization of a greener economy, and lithium-ion batteries are good candidates for such storage. A combined production capacity target of more than \qty{500}{\GWh\per\year} of lithium-ion battery cells was announced in Europe alone \cite{Li-ion_battery_cell_production_Europe_2030}. The present work can be seen as a contribution to help realize this target, by enhancing the understanding of battery thermal effects, effects that are known to be detrimental \cite{Sun2015, An2017,Zhang2020}. Hot spots can arise from internal short circuits and create thermal runaway \cite{Zhang2020}. The local temperature is also important for battery ageing, for so-called cross-over effects and for electrochemical reactions \cite{Sun2015}.

Thermal effects have been studied in detail in the polymer electrolyte fuel cell, including reversible heat effects, see Cai \textit{et al.} for a recent review \cite{Cai2024}. But such heat sources or sinks have been scarcely investigated in lithium-ion batteries. Spitthoff \textit{et al.} \cite{spitthoff_peltier_2021} made a first attempt, focusing on initial stages of cell operation and accounting for the coupling that takes place between fluxes of heat and charge. Recent data from Gullbrekken \textit{et al.} \cite{gullbrekken_coeff} offer an opportunity to also account for the gradients in chemical potential that develop during quasi-stationary cell operation.

To describe coupled transport processes in batteries is a challenge, in particular when it comes to describe the processes taking place at the electrode surfaces. The electrochemical reactivity will vary over the heterogeneous electrode surfaces. Much effort has thus been devoted over the years to find a proper coarse-grained description of all relevant coupled phenomena at the electrode surfaces, see \textit{e.g.} \cite{Dawlaty2023}. The derivation of a continuous description of pore-scale events has been central \cite{analysis_polarization_Li_ion, review_param_database_contin_Li_ion, param_physics_based_models_review}. 

Coupling coefficients were  recently shown to be essential  for the complex electrolyte in the lithium-ion battery \cite{Fong_Transf_Polyelectrolyte, gullbrekken_coeff}.
Nernst-Planck equations, possibly supplied with advection terms, have often been considered to be sufficiently precise for cell modelling, but these equations are now not sufficiently precise; they disregard the coupling between fluxes of heat and mass and charge.

Most available models do not deal with coupled transport phenomena at the electrode surfaces as such, see \textit{e.g.} \cite{Sun2015}. Rules for coupling of chemical and electrical driving forces in the surfaces differ from those of the bulk phases \cite{net_heterogen}. Current continuous descriptions of charge transfer at the electrode-electrolyte interface do not deal with this coupling directly, and can therefore not so easily be used to question common assumptions, for instance that of a constant distribution of reversible heat in the battery. Liu \textit{et al.} \cite{Liu2024} gives a recent overview of thermally coupled battery models. The book of Newman \cite{dfn_model_1994} is a good reference for state-of-the-art modelling in electrochemistry.

In order to describe the electrochemical surface with correct coupling or symmetry properties for transports at the electrode surfaces, we have turned to the theory of nonequilibrium thermodynamics (NET) as written for heterogeneous systems \cite{net_heterogen}. 
Pore-scale events can then be included into a thermodynamic description, by defining the surface as a 2D system of Gibbs surface excess variables. The excess variables are densities of state variables which are integrated over the surface thickness. The surface becomes a 2D autonomous system in the process. The essential theoretical basis was established in the 1970s \cite{bedeaux_bc_net, bedeaux_non_eq_electro}. It was formulated for electrochemical systems much later \cite{net_heterogen}, and has even been used to describe shock waves \cite{Maltby2023}.  This form of NET is particularly suitable when we need to deal with dynamic boundary conditions, \textit{i.e.} when there are sinks and sources of various types in a surface. This is a trademark of electrochemistry, but also of NET for heterogeneous systems, which makes NET suited for this purpose.

We shall therefore build on and expand the NET electrochemical cell model of Spitthoff \textit{et al.} \cite{spitthoff_peltier_2021} in this work. They calculated the temperature profile across a cell accounting for the coupling of heat and charge, including the initial Peltier heats of the electrode surfaces. However, they neglected possible contributions arising from gradients in chemical potential, since they were focusing  on the initial stages of cell operation. Battery cells usually operate in quasi stationary state \cite{korthauer_lithium-ion_2018}, however. With data recently available from Gullbrekken \textit{et al.} \cite{gullbrekken_coeff} there is a possibility to account also for these effects, and improve and extend the work of Spitthoff \textit{et al.} \cite{spitthoff_peltier_2021}.

A possibility is also available with NET to test the battery model used for thermodynamic consistency. Common models of electrochemical systems suffice to use mass-, energy- and momentum balances. NET includes also the entropy balance, which can be used for consistency tests. This will be demonstrated here.  

The aim of the work is thus to obtain an improved model for the lithium-ion battery which takes into account thermal phenomena in a thermodynamically consistent manner. We also show that the description in use has a large impact on the outcome of the analysis. The present work concerns stationary states. It is our hope that the general procedure demonstrated can be used in electrochemical cell modelling at large. A robust program is therefore made available for new users. 

The paper is structured as follows. In Section \ref{sec:cell_layers} we present the model for a layered battery cell, and a choice of components for the particular battery in question. In Section \ref{sec:gov_eq} we give the principles of the theory, and the final equations used, while referring to literature for particular derivations.
The numerical solution procedure is outlined in detail in Section \ref{sec:num_sol}, with the purpose of helping users of the program provided in GitHub. Results and discussion follow in Section \ref{sec:results} for the base case and variations in that. 
All coupling coefficients between heat, mass and charge transport are relevant. This will be pointed out and explained, with emphasis on thermal effects which are new. Particular emphasis will also be made on the possibility to check the model for thermodynamic consistency by using the entropy balance.   

\section{A battery cell of five layers}
\label{sec:cell_layers}

\begin{figure}[htb]
    \centering
    \includegraphics[scale=0.4]{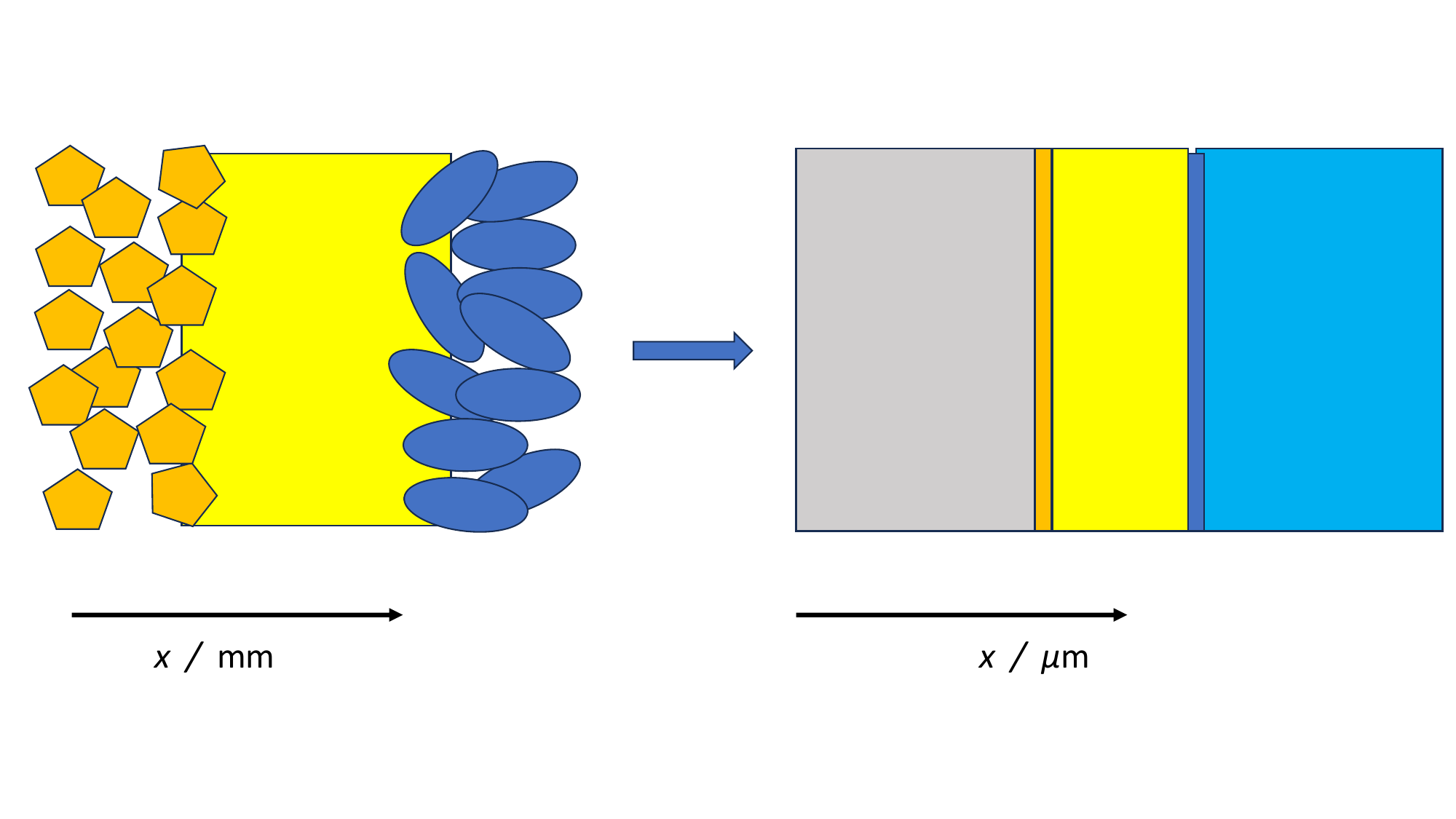}
    \caption{The electrochemical cell on the continuum (right) and on the molecular (left) scale. On the left is illustrated the porous 3D structure of the anode material (orange) in contact with the (yellow) electrolyte and contact between the electrolyte and the cathode material (blue). The interface regions are integrated out to give a 2D representation to the right in the figure.}
\label{fig:simplification_surface}
\end{figure}

The lithium-ion battery was chosen with the dual purpose to produce new information on the temperature profile of this important battery, but also to show how NET can be used to our advantage in battery modelling, \textit{i.e.} to obtain a model compatible with the second law of thermodynamics. When NET is used to describe electrochemical cells, the cell is divided into five (or more) layers, see Figure \ref{fig:simplification_surface}. The figure to the right has three bulk layers; of the cathode (grey color), the anode (blue color) and of the electrolyte mixture (yellow color). The layer thicknesses of these three layers can be described on the continuum level using NET (\unit{\um}-scale). These layers can be considered to consist of 3D homogeneous materials with bulk properties.

To the left in the figure is shown a close-up of the electrolyte with the adjacent electrode surfaces on much smaller scale. The scale is used to measure changes on the nanometer scale. The electrode surface is defined as the region where the material densities vary from the electrolyte-value to zero.
In the construction of Gibbs excess variables, we integrate over these regions. This gives a discrete (2D) description. In this manner we obtain coarse-grained variables; or excess properties of the 
electrode surfaces. The procedure enables us to deal with coupling of fluxes in a 2D-system \cite{net_heterogen}, because the scalar components of the fluxes of mass, heat and charge will couple to the scalar chemical reaction. This is unlike the situation in the bulk phases.

Electrode materials are often porous and multiphase. In a 2D-effective description, the properties become average values over the cross-section of the cell. This way of dealing with surfaces is standardly used in NET. It sets the stage for use of dynamic boundary conditions.
Figure \ref{fig:simplification_surface} shows two surface area elements on the right hand side  (orange and blue stripes). The electrode reactions take place within this region of molecular thickness. The surfaces extend in space so as to include all positions where the electrochemical reaction can take place. By integrating over the surface thickness we obtain a 2D coarse grained system that is represented by a thin vertical line \cite{net_heterogen}, to signify a possible discontinuity.
The notation used for 3D- and 2D phases is taken from ref. \cite{net_heterogen} and illustrated in Figure \ref{fig:notation_thermodynamic_system}. 
We consider transport phenomena perpendicular to these surfaces only, meaning that the system is further reduced to a 2D- and 1D system.

The materials used in the present work are in the essence the same as used by Spitthoff \textit{et al.} \cite{spitthoff_peltier_2021}, cf. the list in Appendix: Material Properties. The bulk materials are:
\begin{itemize}
    \item Negative Electrode:  Lithium intercalated in graphite, \ce{LiC6}
    \item Positive Electrode: Lithium iron phosphate \ce{LiFePO4} (LFP)
    \item Electrolyte: \ce{LiPF6} in 1:1 ratio of EC:DEC
\end{itemize}
The electrode surfaces arise between the negative electrode and the electrolyte and between the electrolyte and the positive electrode (orange/dark blue lines in Fig. \ref{fig:simplification_surface}). The electrolyte in the lithium-ion battery is composed of a lithium salt (\ce{LiPF6}) and often of two organic co-solvent carbonates, here ethylene carbonate (EC) and diethyl carbonate (DEC). The electrolyte may consist of ternary, even quaternary mixtures.

Gullbrekken \textit{et al.} \cite{gullbrekken_coeff} showed that both solvents, EC and DEC, were carried along with the charge transporter in the wall frame of reference, with EC playing a less direct role than DEC. 
Therefore they chose to use the EC frame of reference \cite{gullbrekken_coeff}. In a thermodynamic description of the cell, the surface is the natural frame of reference. 

Gullbrekken \textit{et al.} \cite{gullbrekken_coeff} proposed furthermore to describe the ternary electrolyte as a reacting mixture in equilibrium:
\begin{align*}
    \ce{LiPF6  + 3 DEC <=> \ce{Li+} * 3 DEC + \ce{PF6-}}
\end{align*}
With the EC frame of reference and this equilibrium condition imposed, there are two independent mass variables in the electrolyte, \ce{LiPF6} and DEC. In the presence of thermal, chemical and electrical driving forces, we shall thus find that we need a 4 $\times$ 4 matrix of Onsager coefficients in order to solve the model for the bulk electrolyte.

\section{Governing equations for a battery of five layers}
\label{sec:gov_eq}

Following the procedure of NET we formulate the entropy production for each layer of a cell in the battery mode of operation. The layers are, counting from left to right in Fig. \ref{fig:simplification_surface}; 1) the bulk anode layer, 2) the anode surface, 3) the electrolyte layer, 4) the cathode surface and 5) the bulk cathode layer. The flux-force relations associated with the layers follow from the entropy production. This full derivation was presented for the lithium-ion battery before \cite{net_heterogen, spitthoff_peltier_2021,Kjelstrup2023} and will not be repeated here. We suffice to present the expressions for the entropy production along with the set of governing equations for each layer. The sets of equations, used in the following to determine the temperature- and composition profiles at stationary state, are given in Tables 1-6. We are only concerned with one-dimensional transport, in the direction normal to the surface.

\begin{figure}[htb]
    \centering
    \includegraphics[width=0.8\textwidth]{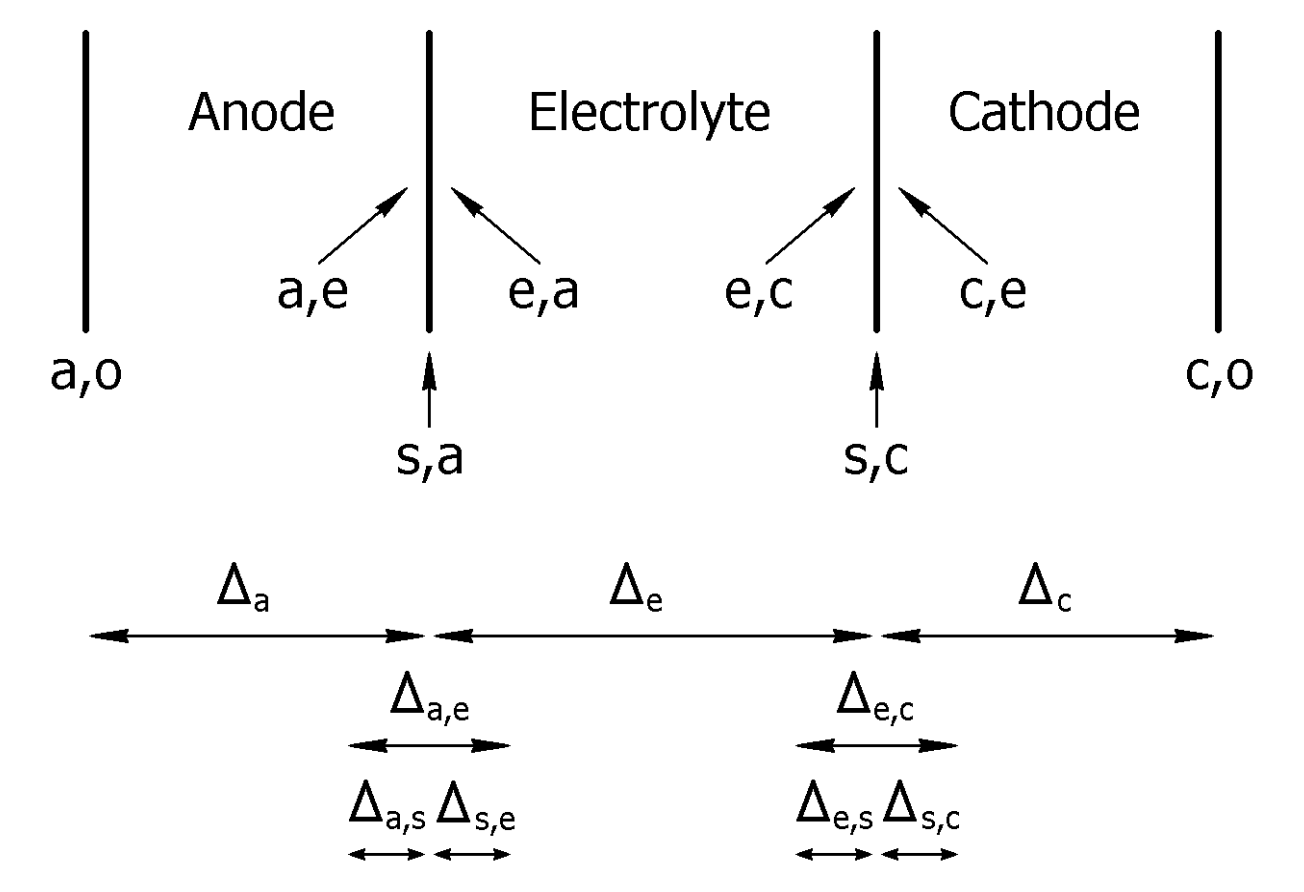}
    \caption{Super- and subscripts used in electrochemical cells. 
    Superscripts a, s and c refer to anode, surface and cathode, respectively. Double super- and subscripts refer to phase location and neighboring phase.}
    \label{fig:notation_thermodynamic_system}
\end{figure}

The notation in use is illustrated in Fig. \ref{fig:notation_thermodynamic_system}. Superscripts a, s and c refer to anode, surface and cathode, respectively. Double super- and subscripts refer to phase location and neighboring phase (layer). So $a,e$ means a position in the anode layer next to the electrolyte.

\subsection{The Anode Bulk Layer}

Equations for the anode bulk phase are presented in Table \ref{table:Governing Equations Electrode a}. The entropy production of the anode layer (anode bulk phase) is given in Eq. \eqref{eq:electrode_sigma_a}. 
In the expression for the entropy production, $\sigma$, \( J'_q \) is the measurable heat flux, \( J_{\text{L}} \) is the flux of the neutral component (metal) L, while \( j \) is the electric current density.  On the left hand side the component is lithium intercalated in graphite, \ce{LiC6}. The terms in the parentheses in Eq. \eqref{eq:electrode_sigma_a} are the conjugate thermodynamic driving forces. The thermal driving force is the gradient in the inverse temperature, $\partial/ (\partial x) (1/T) $.

\vspace{0.5cm}

\begin{table}[H]
    \centering
     \caption{Governing Equations for the Anode Bulk Phase, \ce{LiC6}.}
     \label{table:Governing Equations Electrode a}
    \begin{tabularx}\textwidth{@{}ML@{}}
        \toprule
        \multicolumn{1}{l}{\textbf{Governing Equations Electrode Bulk Phases}} & \multicolumn{1}{l}{\textbf{Ref.}}  \\
        \midrule
        \\
        \frac{\text{d}^2 T}{\text{d}x^2} = \left(c^a_{p,\text{L}} - \frac{\pi^a}{T}\right) \frac{j}{F \lambda^a}  \frac{\text{d}T}{\text{d}x} -  \frac{1}{\kappa^a \lambda^a}  j^2
        & eq:electrode_temperature_a \\
        \\
        \midrule
        \\
        \sigma = J'_q \left( \frac{\text{d}}{\text{d}x}  \frac{1}{T} \right) +  J_{\text{L}} \left( -\frac{1}{T}  \frac{\text{d}\mu _{\text{L}, T}}{\text{d}x}\right) + j \left( -\frac{1}{T}  \frac{\text{d} \phi}{\text{d}x}\right)
        & eq:electrode_sigma_a \\
        \\
        J'_q = -\lambda^a \frac{\text{d}T}{\text{d}x} + \frac{\pi^a}{F}j ; 
        ~~{\pi^a} = -TS_e^*
        & eq:electrode_Jq_a \\
        \\
        J_{\text{L}} = \frac{j}{F} = - D^a_{\text{L}} \frac{\text{d}c_{\text{L}}}{\text{d}x} \Rightarrow \frac{\text{d}c_{\text{L}}}{\text{d}x} = - \frac{j}{D^a_{\text{L}} F}
        & eq:electrode_dc_a \\
        \\
        \frac{\text{d}\mu_{\text{L}, T}}{\text{d}x} = \Gamma^a \frac{RT}{c^a_{\text{L}}} \frac{\text{d}c^a_{\text{L}}}{\text{d}x}
        & eq:electrode_dmu_a \\
        \\
        \frac{\text{d}\phi}{\text{d}x} = -\frac{\pi^a}{TF}\frac{\text{d}T}{\text{d}x} - \frac{j}{\kappa^a}
        & eq:electrode_dphi_a \\
        \\
        \bottomrule
    \end{tabularx}
\end{table}

Each term on the right hand side of $\sigma$ is a product of a flux and its driving force. The graphite lattice is fixed and serves to define the surface frame of reference. The gradient in chemical potential of L, \(\partial \mu_{\textrm{L},T}/\partial x\), is  evaluated at constant $T$. Other variables may be used, but  
all valid sets give the same entropy production \cite{net_heterogen}.

The governing equations contain transport properties. The heat flux has a Fourier type term, where $\lambda^a$ is the thermal conductivity. It also has a Peltier heat contribution, $\pi^a$. For the anode bulk material this is equal to the transported entropy of the electron, $S_e^*$, cf. Eq. \eqref{eq:electrode_Jq_a}. We shall assume that diffusion of lithium takes place  from a reservoir next to the graphite and into the electrode surface. On the average,  $J_{\textrm{L}}=j/F$ for all conditions.  The diffusion coefficient is $D_\text{L}^a$. This diffusion does not contribute to conduction and therefore not to the electric potential gradient in Eq. \eqref{eq:electrode_dphi_a}. This enables us to use Fick's law alone to find the gradient in concentration, $c^a_{\textrm{L}}$. The chemical potential gradient in Eq. \eqref{eq:electrode_dmu_a} is available, as we know the thermodynamic factor of the material, $\Gamma^a$ \cite{gullbrekken_coeff}. The gradient in the electric potential, Eq. \eqref{eq:electrode_dphi_a}, can then be found, once the electric conductivity, $\kappa^a$, is known. The temperature profile is determined from the energy balance, which also contains the heat capacity of L, $c^a_{p,\textrm{L}}$, in addition to other variables.

\subsection{The Anode Surface Layer}

\begin{table}[H]
    \centering
     \caption{Governing Equations for the Anode Surface.}
     \label{table:gov_eqs_anode_surface}
    \begin{tabularx}\textwidth{@{}ML@{}}
        \toprule
        \multicolumn{1}{l}{\textbf{Governing Equations Anode Surface}} & \multicolumn{1}{l}{\textbf{Ref.}} \\
        \midrule
        \\
        \Delta_{a,s}T = \frac{\lambda^a}{\lambda^{s,a}} \frac{\text{d}T}{\text{d}x}\bigg |_{x = x^{a,e}}
        & eq:surface_anode_dTas \\
        \\
        \Delta_{s,e}T = \left[ \left(\lambda^{s,a} - \frac{\pi^{a} j}{F T^{a,e}} \right) \Delta_{a,s}T + \frac{\Pi^{s,a}}{F}j - r^{s,a}j^2 \right] / \left( \frac{\pi^{e} j}{F T^{e,a}} + \lambda^{s,a} \right)
        & eq:surface_anode_dTse \\
        \\
        \frac{\text{d}T}{\text{d}x}\bigg |_{x = x^{e,a}} = \left( \lambda^{s,e} \Delta_{s,e}T + \frac{j}{F} (b_q - \pi^e) \right) / \left( \lambda  -\frac{a_q}{T^2}  \right)
        & eq:surface_anode_dTdx \\
        \\
        \midrule
        \\
        \sigma^{s,a} = J'^{a,e}_q \Delta_{a,s} \left(\frac{1}{T}\right) + J'^{e,a}_q \Delta_{s,e} \left(\frac{1}{T}\right) - j \frac{1}{T^{s,a}} \left[ \Delta_{a,e} \phi + \frac{\Delta_n G^{s,a}}{F} \right]
        & eq:sigma_surface \\
        \\
        \Delta_{a,e} \phi  = - \frac{\pi^a}{T^{a,e}F} \Delta_{a,s}T - \frac{\pi^e}{T^{e,a}F} \Delta_{s,e}T - r^{s,a}j - \frac{\Delta_n G^{s,a}}{F}
        & eq:surface_phi \\
        \\
        \Pi^{s,a} = T \left( \frac{J_s^{e,a}-J_s^{a,e}}{j/F} \right)_{T^{a} = T^s = T^e} = \pi^e - \pi^a - TS^{a}_{\text{L}} = T( - S^*_{\text{Li}^+} + S^*_e  - S_{\text{L}}^{a})
        & eq:peltier_heat_anode_sf \\
        \\
        \eta^{s,a} = r^{s,a} j = \frac{2RT}{F}\ln \left( \frac{j}{j_{0,a}} \right)
        & eq:tafel_equation \\
        \\
        \Delta_nG^{s,a} = -zF\Delta_{i,o}^{eq}\phi & eq:nernst_equation
        \\
        \\
        \bottomrule
    \end{tabularx}
\end{table}

The governing equations of the anode surface are presented in Table \ref{table:gov_eqs_anode_surface}. The equations serve as dynamic boundary conditions for the processes taking place in the battery cell. A simpler name of the equations is anode jump conditions.

A double subscript of the $\Delta$'s in Table \ref{table:gov_eqs_anode_surface} (\textit{e.g. a,e}) means that the difference is taken between position $e$ and $a$. Across the interface, between the bulk anode and the electrolyte, there are possible jumps in the inverse temperature into the surface ($\Delta_{a,s}(1/T)$) and out of the surface ($\Delta_{s,e}(1/T)$), as well as in the electric potential across the surface (\(\Delta_{a,e} \phi\)). The symbols have the same meaning as explained for the bulk anode. In addition, we define the Peltier heat of the anode surface, $\Pi^{s,a}$ in terms of transported entropy of electron, and of Li$^+$ and the metal entropy, cf. Eq. \eqref{eq:peltier_heat_anode_sf}. The overpotential of the anode $\eta^{s,a}$ is obtained from experiments through Eq. \eqref{eq:tafel_equation}. For material properties, see Appendix: Material Properties.

The reaction Gibbs energy is derived from changes in chemical potentials of the neutral components that enter the electrode reaction, \(\Delta_n G^{s,a}\). The chemical and electric forces combine to an effective force \cite{net_heterogen} in the stationary case, when the reaction rate $r$ is given by the current density, $j$, $r=j/zF$.
Here $z=1$ is the number of electrons involved in the electrode reaction and $F$ is Faraday's constant. We assume that 
there are no terms that account for adsorption of components onto the porous material in the excess entropy production, \textit{i.e.} we assume equilibrium for adsorption: \(\Delta_{a,s} \mu_{\text{L}, T} = 0\).

At the electrode surfaces there is always a discontinuity in the charge carrier caused by the electrochemical reaction. The surface may include a double layer, but is chosen to remain electroneutral. The reaction Gibbs energy is given by \cite{net_heterogen}: 
\begin{equation}
    \Delta_n G^s(t) = \sum_{j \in \text{neutral}} \nu_j \mu_j^s
\end{equation}
where the surface means anode surface in this case. 
When the electroneutrality condition is everywhere obeyed, the electric current density is independent of the position. At reversible conditions ($\sigma^s = 0 $), the \textit{Nernst} equation is recovered: 
\begin{equation}
    \Delta_{i,o}^{eq} \phi = - \frac{\Delta_nG^{s,a}}{zF}
    \label{eq:nernst_equation2}
\end{equation}
The electrical force the "open circuit voltage", at ($j=0$) is here balanced by the chemical driving force. The chemical driving force is the reaction Gibbs energy for the neutral components of the electrode reaction. Similar equations are written for the cathode surface below, see below. See also \cite{spitthoff_peltier_2021,Kjelstrup2023,gullbrekken_coeff}. Away from the state of balance, the electric potential jump decreases with $j$ according to the empirical expression of the the overpotential, see Eq. \eqref{eq:tafel_equation}.

\subsection{The Electrolyte Layer}

The entropy production of the electrolyte, cf. Eq. \eqref{eq:electrolyte_sigma} in Table \ref{table:gov_eqs_electrolyte}, has the same number of terms as the electrode layers but with a mass flux added, the flux \(J_{\text{D}}\) of the neutral co-solvent component DEC.
Table \ref{table:gov_eqs_electrolyte} (Eqs. \eqref{eq:electrolyte_temperature}-\eqref{eq:electrolyte_dphi}) gives the governing equations for the electrolyte, expressed by the variables and lumped variables. We observe that all equations are affected by the presence of component D, and new coefficients appear: The transference coefficients for L and D, $t_\textrm{L}$ and $t_\textrm{D}$ and the heat of transfer for L and for D, $q^*_\textrm{L}$ and $q^*_\textrm{D}$. In addition, there are Onsager coefficients for diffusion and interdiffusion, $l_\textrm{LL}$ and $l_\textrm{DD}$ and $l_\textrm{LD}$. These coefficients are now all known \cite{gullbrekken_coeff}.

The equations \eqref{eq:electrolyte_Jq}-\eqref{eq:electrolyte_dphi} are solved for stationary state conditions where the net flux of salt and co-solvent DEC is zero. The condition allows us to eliminate variables in the energy balance. The temperature is formulated using the lumped coefficients that are given in Table \ref{table:coeffs_electrolyte_temp}. These lumped coefficients $a_\textrm{L}, b_\textrm{L}, a_\textrm{D}, b_\textrm{D}$ together with other variables and Eq. \eqref{eq:electrolyte_temperature}, enable us to give an analytical expression for the profiles in the electrolyte.

\begin{table}[H]
    \centering
    \caption{Governing Equations for the Electrolyte Layer.}
    \label{table:gov_eqs_electrolyte}
    \begin{tabularx}\textwidth{@{}ML@{}}
        \toprule
        \multicolumn{1}{l}{\textbf{Governing Equations for the Electrolyte}} & \multicolumn{1}{l}{\textbf{Ref.}}  \\
        \midrule
        \\
        \frac{\text{d}^2 T}{\text{d}x^2} = - j \frac{a _{\phi}}{TF}/\left( \lambda^e - \frac{a_q }{T^2} \right)  \frac{\text{d}T}{\text{d}x} - \frac{b_{\phi}T}{F^2}  j^2 / \left( \lambda - \frac{a_q }{T^2} \right) - \frac{1}{\kappa} j^2/\left( \lambda - \frac{a_q }{T^2} \right)
        & eq:electrolyte_temperature \\
        \\
        \midrule
        \\
        \sigma = J'_q \left( \frac{\text{d}}{\text{d} x}  \frac{1}{T} \right) + J_{\text{L}} \left( -\frac{1}{T}  \frac{\text{d}\mu _{\text{L}, T}}{\text{d}x}\right) + J_{\text{D}} \left( -\frac{1}{T}  \frac{\text{d}\mu _{\text{D}, T}}{\text{d}x}\right) +  j \left( -\frac{1}{T}  \frac{\text{d}\phi}{\text{d}x}\right)
        & eq:electrolyte_sigma \\
        \\
        J'_q = - \lambda \frac{\text{d}T}{\text{d}x} - q_{\text{L}}^*l_{\text{LL}} \frac{\text{d}\mu_{\text{L}, T}}{\text{d}x} \frac{1}{T} - q_{\text{D}}^* l_{\text{DD}} \frac{\text{d}\mu_{\text{D}, T}}{\text{d}x}\frac{1}{T} + \frac{\pi^e}{F}j
        & eq:electrolyte_Jq \\
        \\
        \frac{\text{d}\mu_{\text{L},T}}{\text{d}x} = -\frac{q^*_{\text{L}}}{T}\frac{\text{d}T}{\text{d}x} -\frac{l_{\text{L}\text{D}}}{l_{\text{LL}}}\frac{\text{d}\mu_{\text{D},T}}{\text{d}x} + T\frac{t_{\text{L}}}{l_{\text{LL}}}\frac{j}{F} 
        & eq:electrolyte_dmuL \\
        \\
            \frac{\text{d}\mu_{\text{D},T}}{\text{d}x} = -\frac{q^*_{\text{D}}}{T}\frac{\text{d}T}{\text{d}x} -\frac{l_{\text{D}\text{L}}}{l_{\text{DD}}}\frac{\text{d}\mu_{\text{L},T}}{\text{d}x} + T\frac{t_{\text{D}}}{l_{\text{DD}}}\frac{j}{F}
        & eq:electrolyte_dmuD \\
        \\
        \frac{\text{d}\phi}{\text{d}x} = -\frac{\pi^e}{TF} \frac{\text{d}T}{\text{d}x} - \frac{t_{\text{L}}}{F} \frac{\text{d}\mu_{\text{L}, T}}{\text{d}x} - \frac{t_{\text{D}}}{F} \frac{\text{d}\mu_{\text{D}, T}}{\text{d}x}  - \frac{j}{\kappa}
        & eq:electrolyte_dphi \\
        \\
        \bottomrule
    \end{tabularx}
\end{table}
\begin{table}[H]
    \centering
       \caption{Lumped Coefficients for the Electrolyte used in the Energy Balance to Solve the Temperature Profile.}
       \label{table:coeffs_electrolyte_temp}
    \begin{tabularx}\textwidth{@{}MM@{}}
        \toprule
        \multicolumn{2}{c}{\textbf{Lumped Coefficients for the Electrolyte to Solve the Temperature Profile}} \\
        \midrule
        \\
        a_{\text{L}} = \frac{l_{\text{DD}}(l_{\text{LL}}q_{\text{L}}^* - l_{\text{LD}}q_{\text{D}}^*)}{(l_{\text{LL}}l_{\text{DD}}-l_{\text{LD}}l_{\text{DL}})}
        &
        b_{\text{L}} = \frac{(t_{\text{L}} l_{\text{DD}} - t_{\text{D}} l_{\text{LD}})}{(l_{\text{LL}}l_{\text{DD}}-l_{\text{LD}}l_{\text{DL}})}
        \\ \\
        \midrule
        \\
        a_{\text{D}} = \left(q^*_{\text{D}} - \frac{l_{\text{DL}}}{l_{\text{DD}}} a_{\text{L}}\right)
        &
        b_{\text{D}} = \left( \frac{t_{\text{D}}}{l_{\text{DD}}} - \frac{l_{\text{DL}}}{l_{\text{DD}}} b_{\text{L}} \right)
        \\ \\
        \midrule
        \\
        a_q = (q_{\text{L}}^* l_{\text{LL}} a_{\text{L}} +  q_{\text{D}}^* l_{\text{DD}} a_{\text{D}})
        & 
        b_q = (\pi^e - q_{\text{L}}^* l_{\text{LL}} b_{\text{L}} - q_{\text{D}}^* l_{\text{DD}} b_{\text{D}} )
        \\ \\
        \midrule
        \\
        a_{\phi} = (\pi^e - t_{\text{L}} a_{\text{L}} - t_{\text{D}} a_{\text{D}})
        &
        b_{\phi} = (t_{\text{L}} b_{\text{L}} + t_{\text{D}} b_{\text{D}}) 
        \\ \\
        \bottomrule
    \end{tabularx}
\end{table}

\subsection{The Cathode Surface}

\begin{table}[H]
    \centering
    \caption{Governing Equations for the Cathode Surface.}
    \label{table:Governing Equations Cathode Surface}
    \begin{tabularx}\textwidth{@{}ML@{}}
        \toprule
        \multicolumn{1}{l}{\textbf{Governing Equations Cathode Surface}} & \multicolumn{1}{l}{\textbf{Ref.}} \\
        \midrule
        \\
        \Delta_{e,s}T = \left[\left( \lambda  -\frac{a_q}{T^2}  \right) \frac{\text{d}T}{\text{d}x}\bigg |_{x = x^{e,c}} + \frac{j}{F}(\pi^e - b_q) \right] \frac{1}{\lambda^{s,c}}
        & eq:surface_cathode_dTes \\
        \\
        \Delta_{s,c}T = \left( \left(\lambda^{s,c} - \frac{\pi^{e} j}{F T^{e,c}} \right) \Delta_{e,s}T + \frac{\Pi^{s,c}}{F}j - r^{s,c}j^2 \right) / \left( \frac{\pi^{c} j}{F T^{c,e}} + \lambda^{s,c} \right)
        & eq:surface_cathode_dTsc \\
        \\
        \frac{\text{d}T}{\text{d}x}\bigg |_{x = x^{c,e}} = \frac{\lambda^{s,c}}{\lambda^c}\Delta_{s,c}T
        & eq:surface_cathode_dTdx \\
        \\
        \midrule
        \\
        \sigma^{s,c} = J'^{e,c}_q \Delta_{c,s} \left(\frac{1}{T}\right) + J'^{c,e}_q \Delta_{s,c} \left(\frac{1}{T}\right) - j \frac{1}{T^{s,c}} \left[ \Delta_{e,c} \phi + \frac{\Delta_n G^{s,c}}{F} \right]
        & eq:sigma_surface_cathode \\
        \\
        \Delta_{e,c} \phi  = - \frac{\pi^e}{T^{e,c}F} \Delta_{c,s}T - \frac{\pi^c}{T^{c,e}F} \Delta_{s,c}T - r^{s,c}j - \frac{\Delta_n G^{s,c}}{F} 
        & eq:surface_phi_cathode \\
        \\
        \Pi^{s,c} = T \left( \frac{J_s^{c,e}-J_s^{e,c}}{j/F} \right)_{T^{c} = T^s = T^e} = \pi^e - \pi^c - TS^{c}_{\text{L}}
        & eq:peltier_heat_cathode_sf \\
        \\
        \eta^{s,c} = r^{s,c} j = \frac{2RT}{F}\ln \left( \frac{j}{j_{0,c}} \right)
        & eq:tafel_equation_cathode \\
        \\
        \Delta_nG^{s,c} = -zF\Delta_{i,o}^{eq}\phi & eq:nernst_equation3
        \\
        \\
        \bottomrule
    \end{tabularx}
\end{table}

\subsection{The Cathode Bulk Layer}

On the cathode side, the governing equations have the same form as on the anode side. The difference is that L now means lithium inside \ce{LiFePO4}. Superscripts refer to location $c$. The Peltier heat of the cathode, as computed from the measured Seebeck-coefficient is $\Pi^s,c$, while the overpotential that belongs to this electrode is $\eta^{s,c}$.

\begin{table}[H]
    \centering
    \caption{Governing Equations for the Cathode Bulk Phase, \ce{LiFePO4}.}
    \label{table:Governing Equations Electrode}
    \begin{tabularx}\textwidth{@{}ML@{}}
        \toprule
        \multicolumn{1}{l}{\textbf{Governing Equations Electrode Bulk Phases}} & \multicolumn{1}{l}{\textbf{Ref.}}  \\
        \midrule
        \\
        \frac{\text{d}^2T}{\text{d}x^2} = \left(c^c_{p,\text{L}} - \frac{\pi^c}{T}\right) \frac{j}{F \lambda^c}  \frac{\text{d}T}{\text{d}x} - \frac{1}{\kappa^c \lambda^c}  j^2
        & eq:electrode_temperature_c \\
        \\
        \midrule
        \\
        \sigma = J'_q \left( \frac{\text{d}}{\text{d}x}  \frac{1}{T} \right) +  J_{\text{L}} \left( -\frac{1}{T}  \frac{\text{d}\mu _{\text{L}, T}}{\text{d}x}\right) + j \left( -\frac{1}{T}  \frac{\text{d}\phi}{\text{d}x}\right)
        & eq:electrode_sigma_c \\
        \\
        J'_q = -\lambda^c \frac{\text{d}T}{\text{d}x} + \frac{\pi^c}{F}j
        & eq:electrode_Jq_c \\
        \\
        J_{\text{L}} = \frac{j}{F} = - D_{\text{L}}^c \frac{\text{d}c_{\text{L}}}{\text{d}x} \Rightarrow \frac{\text{d}c^c_{\text{L}}}{\text{d}x} = - \frac{j}{D_{\text{L}}^c F}
        & eq:electrode_dc_c \\
        \\
        \frac{\text{d}\mu_{\text{L}, T}}{\text{d}x} = \Gamma^c \frac{RT}{c_{\text{L}}^c} \frac{\text{d}c_{\text{L}}^c}{\text{d}x}
        & eq:electrode_dmu_c \\
        \\
        \frac{\text{d}\phi}{\text{d}x} = -\frac{\pi^c}{TF}\frac{\text{d}T}{\text{d}x} - \frac{j}{\kappa^c}
        & eq:electrode_dphi_c \\
        \\
        \bottomrule
    \end{tabularx}
\end{table}

\subsection{Summary of theoretical basis}

Tables 1-6 contain all the equations that are used to determine the profiles of temperature, chemical potential and electric potential. The equations for the profiles of interest, together with the entropy production, are what we call the governing equations. They constitute the thermodynamic model that we have chosen. Clearly, the descriptions of the surfaces differ from those of the bulk phases. At the surfaces, we obtain jump conditions, while in the bulk phases, we use a continuum formulation. 

An integration across a heterogeneous surface, for which we know few properties, is then circumvented. 
A continuous variable has been replaced using excess variables of the whole interface layer. A gradient in the heat flux will at the surface then obtain a discontinuity. Surface values may then appear as singularities.
In other words, as we shall demonstrate, this method allows for heat sources or heat sinks inside the surface in terms of Peltier heats. We integrate layer by layer, so that output of the first layer becomes input of the next layer etc. 
With the above formulation, solution of variable profiles are possible.

\section{Numerical solution}
\label{sec:num_sol}

\subsection{Data input and cases solved}

The model was solved with experimental data  given in Appendix: Material Properties. A base case was established with the governing equations solved for these data. The solution was tested for consistency with Spitthoff \textit{et al.} \cite{spitthoff_peltier_2021}, as well as with  the second law of thermodynamics. The model was finally used to examine assumptions which are common in the literature.

\subsection{Simulation tools}

The present lithium-ion battery cell model involves solution of second-order ordinary differential equations (ODEs), as boundary value problems for each of the bulk phases, see previous section.
The equations were implemented and solved in an object-oriented Python program, which is briefly presented in this section. We outline the main elements and provide guidance on how to run a simulation or change material properties.  The entire program is available through the GitHub repository at  the following link:
\url{https://github.com/felixs97/battery_model.git}

The boundary value problems were conveniently solved in ODEs using a numerical approach (a shooting method). The method transformed the problem into an initial value problem by adjusting the initial conditions in an iterative way, until the solution satisfied the specified boundary conditions \cite{numerics_engineers}. The \texttt{scipy} Python library was used to address these challenges; in particular solvers like \texttt{integrate.solve\_ivp}. The root-finding function (\texttt{optimize.fsolve}) was utilized to align the solution with  the particular boundary conditions of \texttt{solve\_ivp}. By default, the \texttt{solve\_ivp} function uses a Runge-Kutta method of order 5(4) (RK5(4)) \cite{2020SciPy-NMeth}.

The \texttt{optimize.fsolve} function provided in the \texttt{scipy} library, is a wrapper around MINIPACK's hybrid algorithm \cite{2020SciPy-NMeth}, useful for finding roots of non-linear functions.  A subroutine which calculates the function is provided. This allowed us to compute the Jacobian using a forward-difference approximation \cite{minpack_hybrd}. 
In our case the function was the difference of the right-hand-side boundary condition and the solution to the set of equations, obtained from an initial guess of the derivative of the temperature. The difference was optimized to be close to zero by changing the initial guess of the derivative of the temperature.

\subsection{The program structure}

The program was divided into four files:
\begin{itemize}
    \item \texttt{params\_LFP.py}
    \item \texttt{params\_sys.py}
    \item \texttt{classes.py}
    \item \texttt{main.py}
\end{itemize}
The first two files, (\texttt{params\_LFP.py} and \texttt{params\_sys.py}), contained all material properties and physical constants. This allowed for easy changes of parameter values. The material properties of the base case scenario were those of Spitthoff \textit{et al.} \cite{spitthoff_peltier_2021}, see Table \ref{table:materials_electrode} in the Appendix.
This provided us with an independent control of their results.

\begin{figure}[ht]
    \centering
    \includegraphics[width=1.2\textwidth]{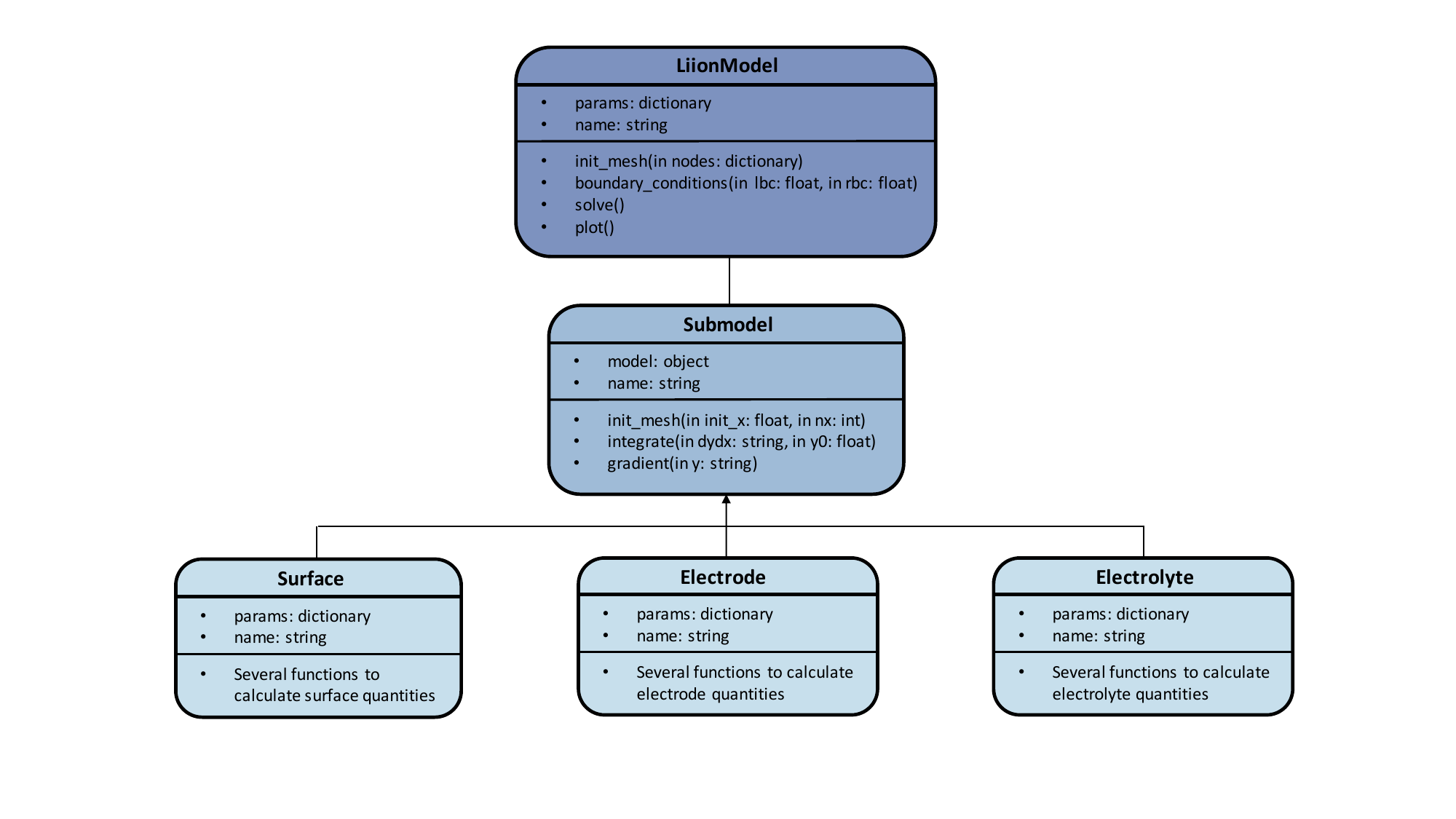}
    \caption{Class Diagram of Cell Model}
    \label{fig:class_diagram}
\end{figure}

The five classes are illustrated in Figure \ref{fig:class_diagram}.
The \texttt{LiionModel} class functions as the overarching container, combining all submodels. Users initiate a simulation by creating an instance of this class, triggering the initialization of the associated submodel.
Upon initializing, a submodel name is assigned to facilitate access to the corresponding parameter set. A variable dictionary (\texttt{vars}) is initialized to store calculated values. Since each submodel is governed by distinct equations, they can be encapsulated in separate classes, with origin in the \texttt{Submodel} base class.
Within each submodel class (\textit{e.g.}, \texttt{Surface}, \texttt{Electrode}, \texttt{Electrolyte}), functions are defined to solve the five sets of governing equations. The entire cell is solved through the \texttt{solve()} method within the \texttt{LiionModel} class.

The solving process begins with the determination of the temperature profile across the cell. This is done by solving the set of equations using the shooting method. The left boundary temperature choice and an initial guess for its derivative are applied and the cell is solved from left to right. The resulting temperature on the right-hand side is then compared to the specified right boundary temperature. An optimization process is applied to ensure agreement with the set boundary condition. Once the temperature profile is known, subsequent quantities can be calculated.

\subsection{Simulation procedure}

In order to execute a simulation, we initiated the \texttt{LiionModel} class and provided required parameters. The initialization process included the creation of a numerical mesh for the solution, achieved by invoking the \texttt{init\_mesh()} function and specifying the number of elements for each bulk phase. The surfaces contributed each with a single element. The boundary temperatures were established by calling the \texttt{boundary\_conditions()} function. Subsequently, the model was solved by invoking the \texttt{solve()} function. To visualise and analyse the results, users can utilize the \texttt{plot()} function. The function provides a comprehensive representation of simulation outcomes. By employing these sequential steps, users can effectively run simulations. Discharging of the cell is modelled by setting a positive current density, while charging of the cell is modelled using a negative current density. An example run and output are shown in the Supplementary Information (SI).

\section{Results and Discussion} 
\label{sec:results}
The results for the base case are presented and discussed first. We continue to show that the case has a model which is consistent with the second law of thermodynamics (Section \ref{sec:consistency}). In Section \ref{sec:assumption}, we investigate the effect of various common assumptions, the effect of changing boundary conditions and the sensitivity of the results to interface material properties.

\subsection{The Base Case}

In the base case scenario we used materials with properties as listed in Appendix: Material Properties. In this scenario, both boundary temperatures were fixed to \qty{290}{\K}. The current density was set to \qty{30}{\A\per\m\squared}, which corresponds to the value obtained by discharging a fully charged cell in \qty{1}{hour} (\qty{1}{C} rate) \cite{spitthoff_peltier_2021}. The lithium concentration in the porous electrodes was derived from the state of charge (SOC) which is approximately \qty{80}{\%} of full intercalation. The material properties are listed with the precision given in the experiment or simulation referred to. They are used in the computations with higher numerical precision than that, to be able to demonstrate principles of the battery cell model. Temperatures may thus be reported with unrealistically high precision. This should be seen as addressing the purpose of method validation.

The results for this scenario are presented and discussed in this section. In all depictions of the result, the anode bulk phase is represented by a blue region on the left-hand side, while the cathode bulk phase is represented by a light red region on the right-hand side. The electrolyte bulk phase is indicated by the white region in between.

\subsubsection{Temperature Profile}
\label{Temperature Profile}

\begin{figure}[H]
    \centering
    \includegraphics[width=0.9\textwidth]{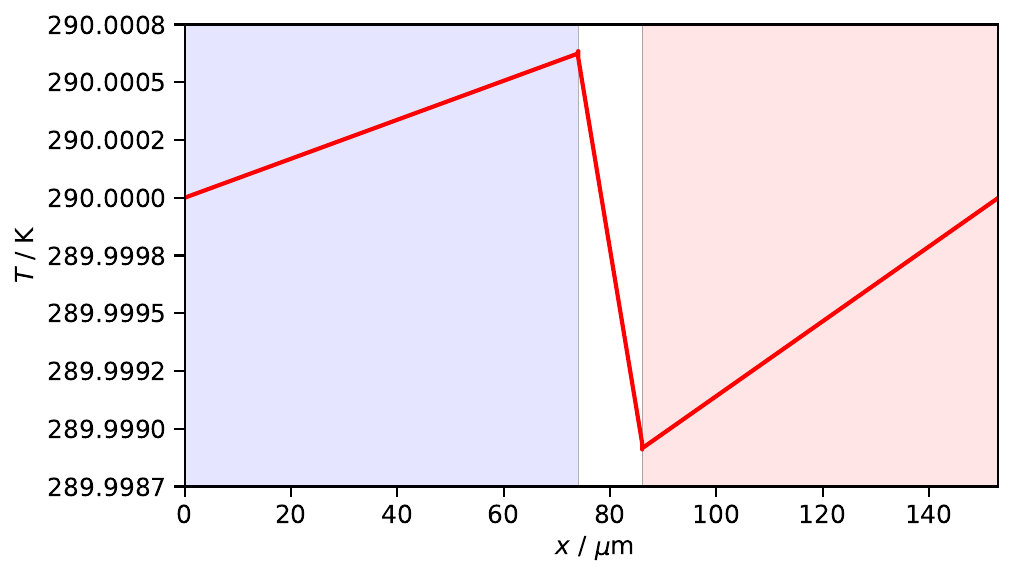}
    \caption{Temperature profile of base case scenario of single battery cell of \ce{C6}/LFP. Anode bulk phase on the left-hand-side in blue, electrolyte bulk phase in white and cathode bulk phase on the right-hand-side in red.}
    \label{fig:temp_profile}
\end{figure}
Figure \ref{fig:temp_profile} illustrates the calculated temperature profile spanning the five layers of the cell.  The temperature varies very little. Slopes are typically 10$^4$ K/m. Such variations were observed before  \cite{spitthoff_peltier_2021} and are characteristic of the variable set. Nevertheless, it illustrates that all of the coupled effects play a role, as we shall see. If they play a role when conditions are mild, they will also play a role away from such conditions.

\begin{figure}[H]
    \centering
    \includegraphics[width=\textwidth]{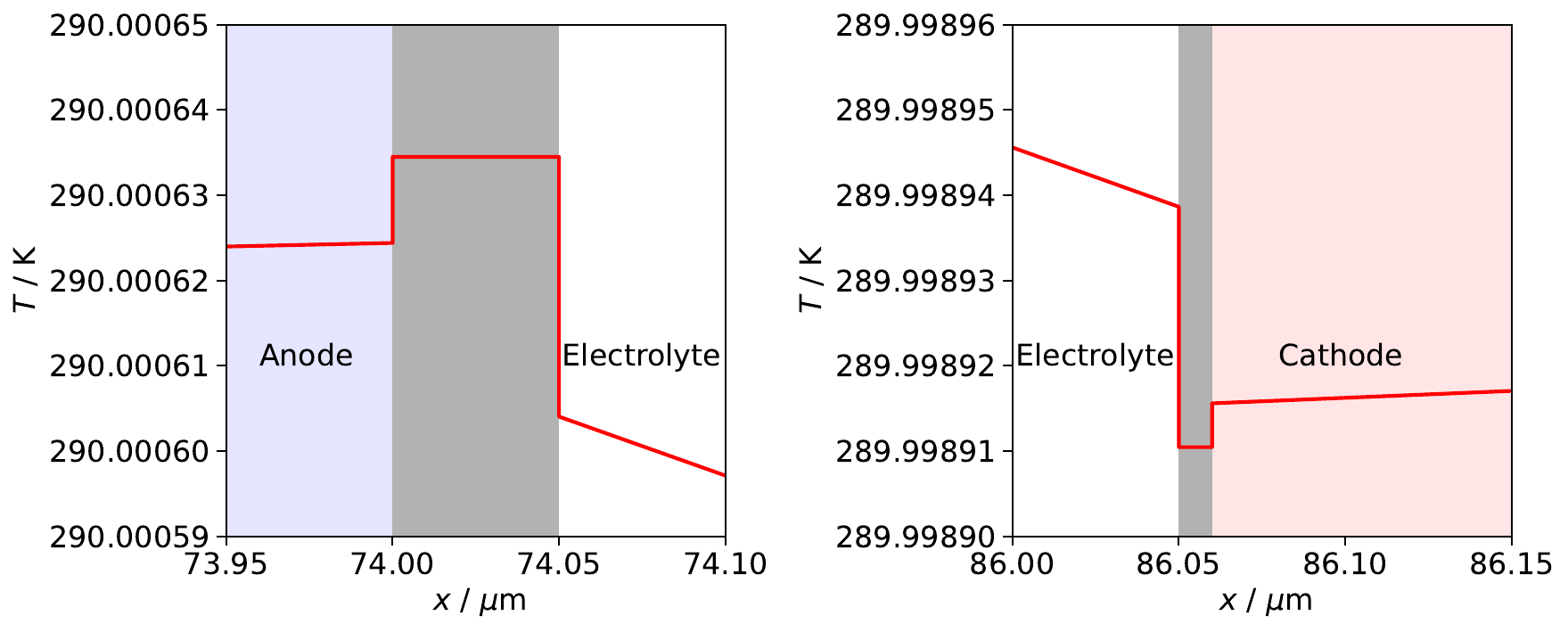}
    \caption{Temperature jumps at the anode and cathode 
    surfaces in the base case scenario of the battery cell \ce{C6}$|$LFP. The anode surface is on the left-hand-side and the cathode surface is on the right-hand-side.}
    \label{fig:temp_profile_surface}
\end{figure}

The surface temperatures are shown in a close-up in Figure \ref{fig:temp_profile_surface}. We see two characteristic properties of the surfaces. First, we see that there is a jump in the temperature across each surface. Secondly, we see that the surface temperature in each electrode distinguishes itself from both adjacent bulk temperatures, it appears as a singularity. These observations follow from the governing equations and are compatible with Peltier heats that are heating the anode and cooling the cathode surface. The observations are all attributable to the respective Peltier heats. Notably, for the \ce{C6}$|$LFP-electrodes, the stationary-state Peltier heats of the two electrodes are of roughly the same magnitude, but opposite sign \qty{-119}{\kJ\per\mol} for the anode, and \qty{122}{\kJ\per\mol} for the cathode, respectively\footnote{Spitthoff et al. \cite{spitthoff_peltier_2021} used in error the opposite sign for the discharge process that we consider.}. Consequently, we observe that the heating and the cooling effects across the cell tend to balance each other, producing a small net entropy change for the cell, as is also observed \cite{gunnarshaug_reviewreversible_2021}.  Moreover, we observe quite linear profiles within the bulk phases, that follow from the fixed boundary temperatures (\qty{290}{\K}), the assumption of constant transport properties and a small current density. A smaller thermal conductivity may create larger local heating or cooling effects. The profile discontinuities are compatible with the discrete description of the processes at the surface, cf. the governing equations of the electrode surfaces. By examining Eq. \eqref{eq:surface_anode_dTas}, we observe that the temperature jump from the anode bulk phase to the surface is equal to the temperature gradient of the bulk phase near the surface scaled by the ratio of the thermal conductivities of the surface and the bulk phase.

\subsubsection{The Measurable Heat Flux}

\begin{figure}[ht]
    \centering
    \includegraphics[width=\textwidth]{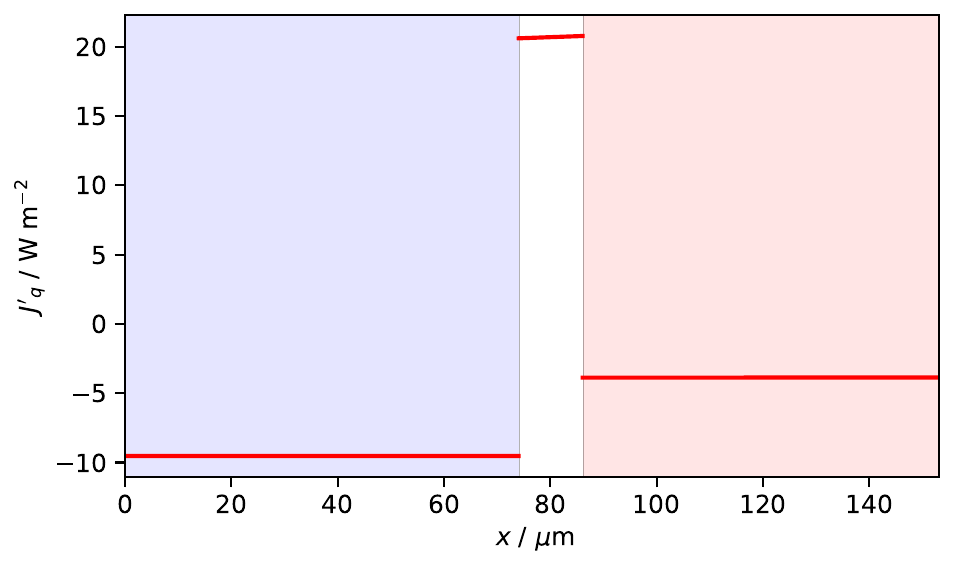}
    \caption{Measurable heat fluxes across the cell of the base case C$_6|$LFP.}
    \label{fig:heat_flux_profile}
\end{figure}

Figure \ref{fig:heat_flux_profile} shows the variation in the measurable heat flux across the cell under the base case scenario. In the bulk phases, the heat flux is constant, compatible with the linear temperature profiles discussed above. The electrolyte shows a very small variation in the slope.  
The change is better seen in Figure \ref{fig:temp_profile_surface}, where the temperature axes of the two figures cover the same range. However, the temperature gradient in the electrolyte deviates slightly between the two plots.

The figure demonstrates again the distinct property of the heat flux; its discontinuity at the interfaces. The enthalpy of lithium changes significantly at both interfaces. When this is added to the (reversible) entropy changes, we obtain work in the outer circuit. The heat fluxes, which contain the Peltier heat, can become large \cite{gunnarshaug_reviewreversible_2021}. In the present case, we noticed even a change in sign in the heat flux. While it is negative in both electrodes (directed to the right-hand side), it is positive in the electrolyte. The finding is compatible with the observed peak in the temperature in the anode surface. The surface temperature is marked by a line, this should not be understood as a variation over the thickness. The surface temperature is a singularity property of the whole layer. Here the surface temperature is higher than elsewhere in the surroundings, meaning that heat leaks to both sides from the anode surface. Similarly, the cathode surface temperature is lower than elsewhere, it is the signature of the surface heat sink.

A technically important observation follows: In the base case scenario, the heat flux in the anode bulk phase is greater in absolute value than the heat flux in the cathode bulk phase. This is explainable by the Peltier heat, which is particular to each electrode. We experience that heat is extracted on the right hand side and added to the left hand side, in order to maintain the boundary temperatures constant on both sides (\qty{290}{\K}).

The heat fluxes in the anode, electrolyte and cathode and their contributions are specified in Table \ref{table:contributions_heat_flux}. The contributions were computed as bulk phase average values.  The table shows the average values (column two), the Fourier term (column three), the contributions from the heat of transfer (columns four and five for the electrolyte only) and the Peltier heats (column six). The Fourier term is always the dominating contribution to the heat flux (compare columns two and three).  Despite the dominance of the Fourier term, we can see significant contributions from coupling terms. In particular, we see that the heat transport induced by the electric current via the Peltier term is not negligible. It can even change sign. In the base case scenario, the Soret contributions from the chemical potential terms in the electrolyte bulk phase almost cancel each other. These terms are significantly smaller than the others. The observations in Table \ref{table:contributions_heat_flux} are unique results for the chosen theoretical description (NET).
The important conclusion is that the existence of small temperature variations do not necessarily mean  that heat fluxes are absent, as may be concluded when Fourier's law is used alone.

\begin{table}[H]
\caption{Contributions (in \unit{\W\per\m\squared} to the measurable heat flux (second column) in the anode, electrolyte and cathode phases.}
\centering
\begin{tabular}{l c c c c c}
\toprule
   Phase & $\begin{aligned}
       J'_q
   \end{aligned}$
   & $\begin{aligned}
       - \lambda \frac{\text{d}T}{\text{d}x}
   \end{aligned}$
   & $\begin{aligned}
       - \frac{q_{\text{L}}^*l_{\text{LL}}}{T} \frac{\text{d}\mu_{\text{L}, T}}{\text{d}x}
   \end{aligned}$ 
    & $ \begin{aligned}
        - \frac{q_{\text{D}}^*l_{\text{DD}}}{T} \frac{\text{d}\mu_{\text{D}, T}}{\text{d}x} 
    \end{aligned}$
    & $  \begin{aligned}
        \frac{\pi}{F}j 
    \end{aligned}$  \\
    \midrule
Electrolyte & \num{21.2} & \num{28.3} & \num{1.6} &  \num{-1.0} & \num{-7.7} \\
Anode & \num{-15.4} & \num{-14.9} & - &  - & \num{-0.5} \\
Cathode & \num{-2.4} & \num{-3.7}  & - & - & \num{1.3} \\
    \bottomrule
\end{tabular}
\label{table:contributions_heat_flux}
\end{table}

\subsubsection{The electric potential profile}

The electric potential profile across the cell reflects the battery voltage in the base case scenario. The voltage was calculated as the difference between the cathode and anode potentials. During battery operation, the cell potential was approximately \qty{3.22}{\V}, indicating a reduction of about \qty{0.33}{\V} of the open circuit voltage (OCV) of \qty{3.55}{\V}. The plots in Fig. S1 in the SI illustrate the jumps at the interfaces at OCV conditions ($j=0$), and in the base case scenario (labelled In operation). The OCV value is reduced primarily because there are overpotentials resulting from the electrochemical reactions (see Eqs. \eqref{eq:tafel_equation} and \eqref{eq:surface_phi}).
\begin{align*}
    \eta^{s,a} + \eta^{s,c} = 0.32 \text{ V}
\end{align*}

In the bulk of the electrode materials, examination of the potential gradient in Eq. \eqref{eq:electrode_dphi_a} revealed contributions from both thermal polarization as well as ohmic terms. The ohmic terms dominated the overall potential gradient in both electrodes, but the integrated potential losses across the electrodes were altogether negligible, due to their high electric conductivities. The OCV reductions added around \qty{3e-4}{\V}.
\begin{equation}
    \frac{\text{d}\phi}{\text{d}x}  = -\frac{\pi}{F}\frac{\text{d}T}{\text{d}x}\frac{1}{T} - \frac{j}{\kappa}
    \tag{\ref{eq:electrode_dphi_a}}
\end{equation}

In the electrolyte the contributions  arose from concentration polarization \cite{pot_MRI_superconc_elec}, thermal polarization and ohmic loss, as indicated in Eq. \eqref{eq:electrolyte_dphi}. 
\begin{equation}
    \frac{\text{d}\phi}{\text{d}x}  = -\frac{\pi}{F}  \frac{\text{d}T}{\text{d}x}\frac{1}{T} - \frac{t_{\text{L}}}{F} \frac{\text{d}\mu_{\text{L}, T}}{\text{d}x} - \frac{t_{\text{D}}}{F} \frac{\text{d}\mu_{\text{D}, T}}{\text{d}x} - \frac{j}{\kappa}
    \tag{\ref{eq:electrolyte_dphi}}
\end{equation}
By examining individual terms (Figure S2), we found that the ohmic contribution was about a third of the total potential gradient. More than half could be attributed to concentration polarization of salt. The co-solvent gave a smaller contribution, but still greater than \qty{10}{\%} \cite{gullbrekken_coeff, cosolvent_imbalance_overpot}. Notably, thermal polarization was here negligible. This analysis underscores the significance of the coupling coefficients to the overall potential drop.

\subsubsection{The concentration profile}

Movements of all components before the stationary state is reached, explain that concentration polarizations arise. We observed gradients in lithium in the electrodes, of the \ce{LiPF6} salt and solvent DEC in the electrolyte. At stationary state, the lithium flux in the electrode was constant, and counted positive as $j/F$ when positive charge is transferred in the electrolyte from left to right. Other component fluxes were then zero. A constant diffusion coefficient resulted in linear concentration profiles.


The fluxes were measured relative to the EC frame of reference. Delving into local effects, we observed accumulation of salt at the anode side, represented by a concentration drop of approximately \qty{100}{\mol\per\m\cubed} across the electrolyte. Simultaneously, the concentration of DEC increased by the same magnitude at the cathode side, resulting in accumulation of DEC at the cathode. The concentrations of the components remained constant everywhere, as we assume that no net component entered or left the cell. 

Equation \eqref{eq:electrolyte_dmuL} 
shows the contributions to the chemical potentials.
\begin{align}
    \frac{\text{d}\mu_{i,T}}{\text{d}x}=-\frac{q^*_i}{T}\frac{\text{d}T}{\text{d}x}-\frac{l_{ij}}{l_{ii}}\frac{\text{d}\mu_{j,T}}{\text{d}x}+T\frac{t_i}{l_{ii}}\frac{j}{F}
    \tag{\ref{eq:electrolyte_dmuL}}
\end{align}
with \(i\) and \(j\) used as placeholders for components, L or D. We estimated the gradients using the transference coefficients of Gullbrekken \textit{et al.} \cite{gullbrekken_coeff}.

The transference coefficient of the DEC component was reported to \num{0.90}. This means, with the passage of 1 faraday of electric charge through the electrolyte, that \qty{0.9}{\mol} of DEC is transported along relative to EC.  Conversely, for the salt component (\(t_{\text{L}} = -0.97\)) the measurement indicated an accumulation of salt on the anode side. The transports interact with each other, as the chemical potential gradient of DEC significantly contributes to the chemical potential gradient of the salt via the coupling coefficients, and vice versa. The temperature gradient contributes in general positively to the electric potential gradient, though it appears negligible in this context, see Table \ref{table:contributions_chemical_potential}.

\begin{table}[H]
\centering
\begin{tabular}{c c c c c}
    \toprule
    $i,j$ &
        $\begin{aligned}
            \frac{\text{d}\mu_{i,T}}{\text{d}x}
        \end{aligned}$
        &
    $\begin{aligned}
        -\frac{q^*_i}{T}\frac{\text{d}T}{\text{d}x}
    \end{aligned}$ 
        &
    $ \begin{aligned}
         -\frac{l_{ij}}{l_{ii}}\frac{\text{d}\mu_{j,T}}{\text{d}x}
    \end{aligned}$ 
        &
    $ \begin{aligned}
        T\frac{t_i}{l_{ii}}\frac{j}{F}
    \end{aligned}$ \\
    \midrule
    L,D & \num{-27.3e6} &  \num{786} & \num{-19.1e6} & \num{-8.15e6} \\
    \midrule
    D,L & \num{6.26e6} & \num{147} & \num{5.74e6} & \num{0.521e6} \\
    \bottomrule
\end{tabular}
\caption{Contributions to the chemical potential gradient in the electrolyte (in \unit{\J\per\mol\per\m}). Results are given as mean values.}
\label{table:contributions_chemical_potential}
\end{table}

\subsubsection{The cumulative entropy production}

\begin{figure}[H]
    \centering
    \includegraphics[width=\textwidth]{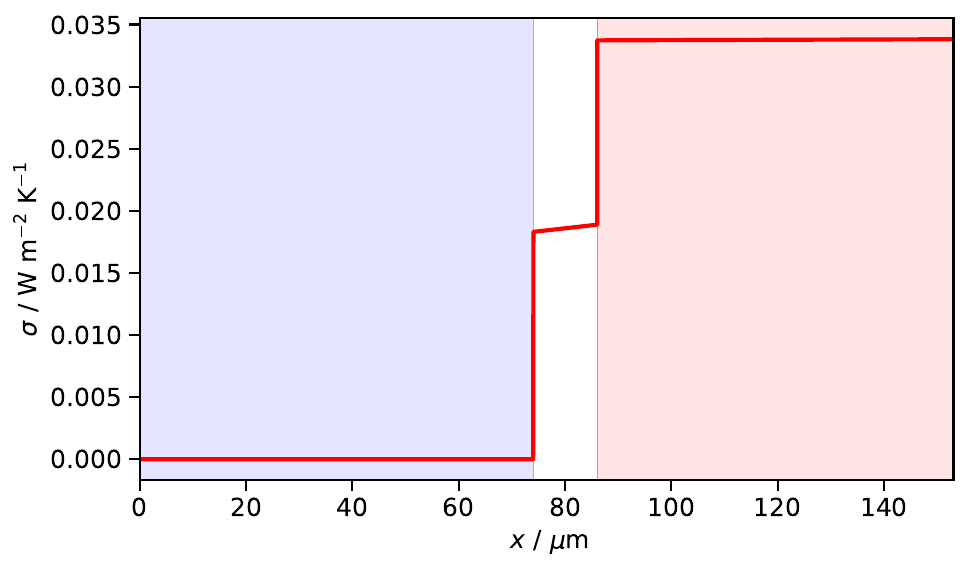}
    \caption{Cumulative entropy production in \ce{C6}$|$LFP.}
    \label{fig:entropy_profile}
\end{figure}

The entropy production was computed in all volume and area elements. The cumulative value is depicted in Figure \ref{fig:entropy_profile}, and contributions to the bulk phases are shown in Table \ref{table:contributions_entropy}. We see that losses which stem from the electrode surfaces dominate, while losses within the bulk of the electrodes are negligible. Among the contributions to Eq. \eqref{eq:sigma_surface}, \qty{99}{\%} of the total entropy production stems from the last term. This term accounts for the difference between the actual electrode potential \(\Delta_{a,e}\phi\) and the reversible cell potential \(\Delta_n G^{s,a}/F\). The same holds true for the cathode surface.
\begin{equation}
    \sigma^{s,a} = J'^{a,e}_q \Delta_{a,s} \left(\frac{1}{T}\right) + J'^{e,a}_q \Delta_{s,e} \left(\frac{1}{T}\right) - j \frac{1}{T^{s,a}} \left[ \Delta_{a,e} \phi + \frac{\Delta_n G^{s,a}}{F} \right]
    \tag{\ref{eq:sigma_surface}}
\end{equation}
In the electrolyte bulk phase, the predominant contribution to entropy production arises from the term that contains the electric current in Eq. \eqref{eq:electrolyte_sigma}. Due to the considerably lower electrical conductivity of the electrolyte compared to that of the electrode bulk phases, the resistivity is here significantly higher, resulting in more losses. 
This is of course highly case-dependent, see next section. 

Table \ref{table:contributions_entropy} shows the contributions to the local entropy production in the bulk phases. In the anode bulk phase, over a third of the local entropy production arises from the heat transport term, while a minimal portion is attributed to the electric current term. Conversely, in the cathode bulk phase, the situation is reversed. In both cases, more than \qty{50}{\%} of the contributions are associated with the diffusion term. The electrolyte dissipation is by far the most important of the bulk phases. 

\begin{table}[H]
\centering
\caption{Contributions to the local entropy production in the bulk phases (\unit{\W\per\m\cubed\per\K}), represented as mean values.}
\label{table:contributions_entropy}
\begin{tabular}{l c c c c}
\toprule
    Phase & $\sigma$ 
    &
    $ \begin{aligned}
        J'_q \left( \frac{\text{d}}{\text{d}x}  \frac{1}{T} \right)
    \end{aligned}$
    &
    $\begin{aligned}
            J_{\text{L}} \left( -\frac{1}{T}  \frac{\text{d}\mu _{\text{L}, T}}{\text{d}x}\right)
        \end{aligned}$
    &
    $\begin{aligned}
        j \left( -\frac{1}{T}  \frac{\text{d}\phi}{\text{d}x}\right)
    \end{aligned}$
    \\
    \midrule
    Anode & \num{0.056} & \num{0.021} & \num{0.033} &  \num{0.012} \\
    Cathode & \num{1.39} & \num{0.03} & \num{0.90} & \num{0.46} \\
    Electrolyte & \num{47.93} & \num{0.03} & - & \num{47.90} \\
    \bottomrule
\end{tabular}
\end{table}

\subsection{The entropy balance: A measure of consistency}
\label{sec:consistency}


Within the framework of NET there is an option to assess a model for thermodynamic consistency. Such a check has been used for fuel cells \cite{net_heterogen}, but we are not aware of any systematic use. The  validation process entails computing the entropy production in two ways, which must give the same result. The method applies to steady state conditions. At steady state there is no change in entropy density, \(s\), with time, and the entropy balance becomes: 
\begin{equation}
    \int_i^o \sigma \ \text{d}x  = J^o_s - J^i_s
    \label{eq:consistency_check}
\end{equation}
Superscript \(i\) and \(o\) describe here the incoming and outgoing flux of entropy, respectively. The expression can be used for single layers, but also for the total cell.
The entropy flux with possible transport of Li on out of the bulk cathode (the $o$-side) is:
\begin{equation}
    J_s^o = \frac{J_q^{'o}}{T^o} + J_{\text{Li}}^o S_{\text{Li}}^o
    \label{eq:entropyflux}
\end{equation} 
and similarly for the $i$-side.
Each layer-model of a battery lends itself naturally to an examination of consistency with the second law of thermodynamics, using Eq. \eqref{eq:entropyflux} and a corresponding equation on the $i$-side, on the right-hand side of Eq. \eqref{eq:consistency_check} comparing it to the integral in Eq. \eqref{eq:consistency_check} for the layer thickness.

In the following, we show that the identity is obeyed for the single electrolyte layer and the single electrode surfaces (Section \ref{sec:entropy_electrolyte} and \ref{sec:entropy_surfaces}). These are the locations in the total cell where almost all the dissipation takes place. 
For the remaining layers, the bulk parts of the anode and the cathode, we shall use the identity to compute an unknown property, namely the molar entropy of Li at the boundary of the layer, see Section \ref{sec:entropy_electrodes}. Thereby, we shall demonstrate an additional usefulness of Eq. \eqref{eq:consistency_check}.

\subsubsection{Entropy production of the single electrolyte layer}
\label{sec:entropy_electrolyte}

The entropy production of the electrolyte layer was computed from the difference between the entropy fluxes at the layer boundaries, as well as from the integral over the electrolyte of the local value of a continuous description of the electrolyte, as described above. In a stationary state, there is no net flux of neither solvent nor salt that carry molar entropy. Consequently, the entropy flux in Eq. \eqref{eq:entropyflux} is determined by the heat flux alone. The expression for the entropy production was given in Table \ref{table:gov_eqs_electrolyte}.

The results for the electrolyte are presented in the top of Table \ref{table:consistency_all_layers}. The numbers for the two calculations are the same within the significant digits that are given. We obtained \qty{5.75e-2}{\W\per\m\squared\per\K} when both external temperatures were 290 K, and \qty{66.12e-2}{\W\per\m\squared\per\K} when there was a difference of 1 K between the two boundary temperatures. All temperatures refer to the boundaries of the total cell, while the temperatures bounding the electrolyte layer are computed in the model.

\newpage

\renewcommand{\arraystretch}{1.0}
\begin{table}[hbt]
\centering
\caption{Test of consistency using the entropy balance for the anode surface, the electrolyte, and the cathode surface. For the bulk anode and cathode materials, the entropy balance is used to compute the molar entropy of lithium. Values for the total cell are computed by integrating over all layers, comparing to the net entropy flux out of the total cell. Numbers are given with the precision of the computation in  \unit{\W\per\cm\squared\per\K}. Results for two different boundary conditions are given, equal boundary temperatures (left hand side column of numbers) and a boundary temperature difference of \qty{1}{\K} (right hand side column of numbers).}
\begin{tabular}{l c c}
    \toprule
    \textbf{Cell part} & {\boldmath \qty{290}{\K}} \textbf{both} & {\boldmath \qty{290}{\K}} \textbf{left -} {\boldmath \qty{291}{\K}} \textbf{right} \\
    \midrule
         \textbf{Electrolyte} & &  \\
     Entropy flux difference & 5.75 & 66.12 \\
     Entropy production      & 5.75 & 66.12 \\ 
     \midrule
    \textbf{Anode surface} & &  \\
     Entropy flux difference & 183.0 & 185.1 \\
     Entropy production      & 183.0 & 185.1 \\
     \midrule 
         \textbf{Cathode surface} & &  \\
     Entropy flux difference & 148.4 & 150.2 \\
     Entropy production      & 148.4 & 150.2 \\ 
     \midrule
         \textbf{Total cell} & &  \\
     Entropy flux difference & 338.2 & 686.2 \\
     Entropy production      & 338.2 & 686.2 \\
    \bottomrule
\end{tabular}
\label{table:consistency_all_layers}
\end{table}

\subsubsection{The entropy production of the anode and cathode surfaces}
\label{sec:entropy_surfaces}

The consistency test was next carried out for the surfaces. The incoming and outgoing measurable heat flux and the entropy change due to transport of lithium give the following difference in entropy fluxes at the anode surface:
\begin{equation}
    J^{e,a}_s - J^{a,e}_s = \frac{J'^{e,a}_q}{T^{e,a}} - \frac{J'^{a,e}_q}{T^{a,e}} - J^{a,e}_{\text{L}}S^{a,e}_{\text{L}}
\end{equation}
Here \(J^{a,e}_{\text{L}} = j/F\) is the flux of lithium and \(S^{a,e}_{\text{L}}\) is the partial molar entropy of L in the anode bulk phase close to the surface.   

The partial molar entropy of Li was determined, using the definition of the Peltier heat of the electrode surface, Eq. \eqref{eq:peltier_heat_anode_sf}. We used the experimental value of 
\(\Pi^{s,a}\) and estimates of \(\pi^e\) and \(\pi^a\) as reported from experiments \cite{gunnarshaug_reviewreversible_2021} (cf. Appendix: Material Properties) to obtain: 
\begin{equation}
    S^{a,e}_{\text{L}} = \frac{1}{T^{a,e}} \left(\pi^e - \pi^a - \Pi^{s,a}\right)
\end{equation} 
Likewise, at the cathode surface, there is a flux of lithium leaving the surface, due to the electrochemical reaction. The difference in the entropy fluxes was expressed as: 
\begin{equation}
    J^{c,e}_s - J^{e,c}_s = \frac{J'^{c,e}_q}{T^{c,e}} - \frac{J'^{e,c}_q}{T^{e,c}} + J^{c,e}_{\text{L}}S^{c,e}_{\text{L}}
\end{equation}
The partial molar entropy of the cathode bulk phase close to the surface \( S^{c,e}_{\text{L}} \) was determined in a similar way as the anode value. The definition of the Peltier heat, Eq. \eqref{eq:peltier_heat_cathode_sf}, gives:
\begin{equation}
    S^{c,e}_{\text{L}} = \frac{1}{T^{c,e}} \left(\Pi^{s,c} + \pi^e - \pi^c \right)
\end{equation}
We recall that the Peltier heat is computed from the experimental value of the Seebeck coefficient using their Onsager relation \cite{gunnarshaug_reviewreversible_2021}. 

The results computed for the partial molar entropy of lithium from these equations gave \qty{275}{\J\per\mol\per\K} for the anode surface and \qty{321}{\J\per\mol\per\K} for the cathode surface. These values are unusually large. The experiment may, however, contain contributions from phase transitions \cite{gunnarshaug_2023}. This point may need further clarification. 

Using these values, we were able to perform the consistency test, with the outcome as given in Table \ref{table:consistency_all_layers}. The model in use for the two surfaces was consistent, with an entropy production of \num{183.0} and \qty{148.4}{\W\per\cm\squared\per\K}, for the anode surface and cathode surface, respectively, for the equal boundary temperature condition. The numbers are about two orders of magnitude larger than the value for the electrolyte. It is expected that most of the dissipation takes place at the surfaces. The values do not vary much with these boundary temperatures, a 1-2 \% change was seen.

\subsubsection{The entropy production of the bulk electrode layers}
\label{sec:entropy_electrodes}

A consistency check of the model for the bulk of the electrodes could not be carried out, for lack of information of the molar entropy in these layers.
The model for the bulk electrodes assumes that the lithium storage capacity in the anode is infinitely large, and that no lithium depletes/accumulates in the anode/cathode. We anticipate a variation in the partial molar entropy with \(x\) \cite{gunnarshaug_2023}.
The difference in entropy fluxes for the anode bulk phase amounted to:
\begin{equation}
    J^{a,e}_s - J^{a,o}_s = \frac{J'^{a,e}_q}{T^{a,e}} + J^{a,e}_{\text{L}}S^{a,e}_{\text{L}} - \frac{J'^{a,o}_q}{T^{a,o}} - J^{a,o}_{\text{L}}S^{a,o}_{\text{L}}
\end{equation}
The molar flux of lithium remained constant throughout the electrode. It is given by the electric current density \( J^{a,e}_{\text{L}} = J^{a,o}_{\text{L}} = j/F\). In the preceding subsection, we estimated a value for the partial molar entropy of lithium close to the anode surface  \(S^{a,e}_{\text{L}}\) using measured Peltier heats.  We now utilize the entropy balance to estimate the partial molar entropy of lithium on the left-hand side of the anode bulk phase, \(S^{a,o}_{\text{L}}\). From Eq. \eqref{eq:consistency_check}:
\begin{equation}
    \begin{aligned}
        \int^{a,e}_{a,o} \sigma \ \text{d}x &= \frac{J'^{a,e}_q}{T^{a,e}} - \frac{J'^{a,o}_q}{T^{a,o}} + \frac{j}{F}(S^{a,e}_{\text{L}} - S^{a,o}_{\text{L}})
        \\
        \Rightarrow 
        S^{a,o}_{\text{L}} &= \left( \frac{J'^{a,e}_q}{T^{a,e}} - \frac{J'^{a,o}_q}{T^{a,o}} - \int^{a,e}_{a,o} \sigma \ \text{d}x \right)\frac{F}{j} + S^{a,e}_{\text{L}} 
    \end{aligned}
\end{equation}
Following this procedure, we ensure that the outcome for the anode bulk phase obeyed the entropy balance.

The same approach was next applied to the cathode bulk phase. We used the partial molar entropy close to the cathode surface, \(S^{c,e}_{\text{L}}\), from the previous subsection, and obtained:
\begin{equation}
    \begin{aligned}
        \int^{c,o}_{c,e} \sigma \ \text{d}x &= J^{c,o}_s - J^{c,e}_s = \frac{J'^{c,o}_q}{T^{c,o}} + J^{c,o}_{\text{L}}S^{c,o}_{\text{L}} - \frac{J'^{c,e}_q}{T^{c,e}} - J^{c,e}_{\text{L}}S^{c,e}_{\text{L}}
        \\
        &= \frac{J'^{c,o}_q}{T^{c,o}} - \frac{J'^{c,e}_q}{T^{c,e}} + \frac{j}{F}(S^{c,o}_{\text{L}}-S^{c,e}_{\text{L}})
        \\
        \Rightarrow
        S^{c,o}_{\text{L}} &= \left(\int^{c,o}_{c,e} \sigma \ \text{d}x + \frac{J'^{c,e}_q}{T^{c,e}} - \frac{J'^{c,o}_q}{T^{c,o}} \right) \frac{F}{j} + S^{c,e}_{\text{L}}
    \end{aligned}
\end{equation}

The model for transport in the bulk electrode is now by design  in agreement with the second law of thermodynamics, cf. Table \ref{table:consistency_all_layers}. The results give a negligible value for the entropy production of these layers when the boundary temperatures are similar. When the temperatures on both sides are \qty{290}{\K}, we obtain an entropy production of \num{0.0262} for the bulk anode and \qty{0.918}{\W\per\cm\squared\per\K} for the bulk cathode. However, when the boundary temperatures are different (right-most column in Table \ref{table:consistency_all_layers}), the heat fluxes increase significantly resulting in entropy production of \num{68.9} and \qty{215.8}{\W\per\cm\squared\per\K} in the bulk anode and cathode layers, respectively.


The value obtained for the entropy production of the total cell can now be computed. When both external temperatures are \qty{290}{\K}, we obtain \qty{338.2}{\W\per\cm\squared} with both ways of calculation. The electrode bulk values are needed to make a perfect agreement.

The entropy production of the total cell represents a dissipation of \qty{9.8}{\W\per\m\squared}, a small value typical for this type of batteries. The value may change much, \textit{i.e.} when the current density is increased, but this is beyond this work to address.

A thermodynamic model is never unique, in the sense that a single model may not be the only model that has a property. Here we have shown how the entropy balance can be used in electrochemistry for two purposes. One purpose can be to check a thermodynamic model for consistency. The entropy balance can also be used as an independent equation to find unknown properties.

\subsection{Checking assumptions common in the literature}
\label{sec:assumption}

\begin{figure}[htb]
    \centering
    \includegraphics[width=\textwidth]{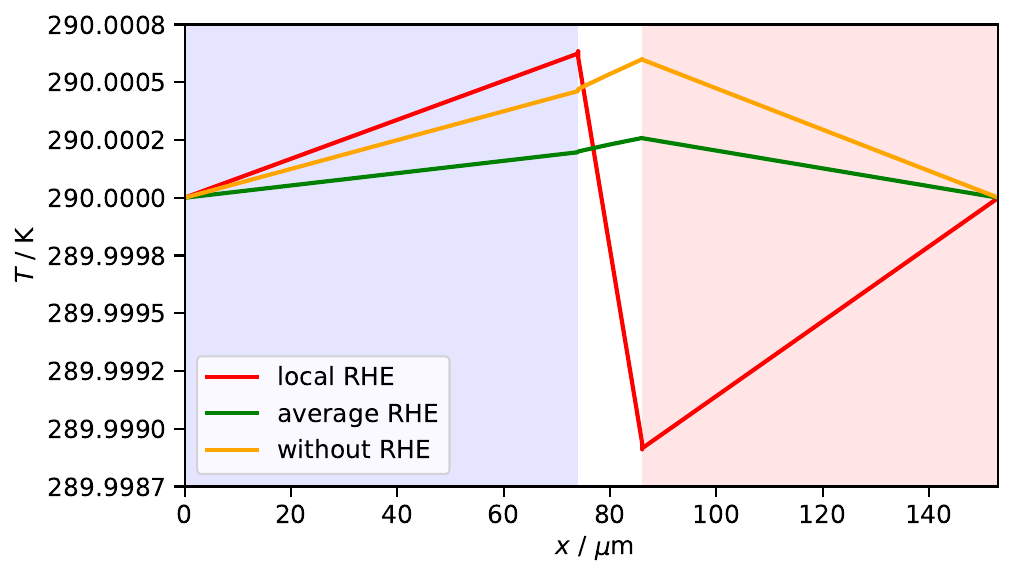}
    \caption{Temperature profile of single battery cell \ce{C6}$|$LFP for different assumptions about the reversible heat effects. The thermodynamically consistent model is pictured by a red line. The green line illustrates a model where the total reversible heat effect RHE is distributed uniformly across the cell, while the yellow line shows results where RHE is neglected.}
    \label{fig:temp_profile_rhe}
\end{figure}

When the entropy balance is everywhere obeyed, we know that a model is thermodynamically consistent. We are then also well set to explore cases outside the base case scenario, and test the model's sensitivity to changes in thermal properties. We have chosen to delve into the impact of changing thermal properties as these are frequently neglected or treated by average procedures. We have therefore explored scenarios, where the total reversible heat effect (the cell entropy change) is distributed uniformly across the cell. This is called the average Reversible Heat scenario (aRH) and is commonly used in literature \cite{thermal_cond_internal_temp_prof, invest_li_ion_batt_echem, DAMAY201537, DU2015327}. We also studied the case when it is neglected all together, the neglected Reversible Heat scenario (nRH). The effects on temperature profiles and measurable heat fluxes, as explored for the aRH and nRH scenarios, are shown in Fig. \ref{fig:temp_profile_rhe}. 
In the base case scenario, we have used electrode specific Peltier heats derived from measurements \cite{gunnarshaug_reviewreversible_2021}.
Notably, we observe heating at the anode surface and cooling at the cathode surface. In the aRH scenario, the total heat effect, $T \Delta S$, is evenly distributed between both surfaces. In the third case, reversible heat effects are neglected altogether, and contributions arise solely from ohmic heating and concentration polarization.

Examining the heat fluxes of the cell, shown in Table \ref{table:heat_generation_aRH}, we note a significant underestimation of the heat that needs to be dissipated to maintain constant boundary temperatures when we neglect the RH. Whether we assume local or average aRH does not affect the outcome, because the overall change in \(T \Delta S\) is the same in both cases. However, we observe a marked disparity between local and average RH in the temperature profile. This underscores the importance of the value of the heat flux. They influence the local temperatures. It is a rather small effect in the present case which contains one unit cell only. It may grow several degrees above \qty{290}{\K} in a cell stack \cite{spitthoff_peltier_2021}. To incorporate reversible heat effects in a proper manner is thus essential for accurate temperature modelling of electrochemical cells. Specifically, to assume that heat is evenly distributed (aRH) may lead to underestimation of local temperature peaks. The approach to steady state from a constant temperature scenario does not alter conclusion. To use Fourier's law for heat conduction alone, leads to other errors.
\begin{table}[ht]
\centering
\caption{Measurable heat flux requiring dissipation to uphold consistent boundary temperatures. All units are in \unit{\W\per\m\squared}}
\begin{tabular}{l l l l}
    \toprule
    Heat flux & Base case & Average reversible  & No reversible \\ 
    difference & - & heat effect & heat effect \\
    \midrule
    $J'^{c,o}_q - J'^{a,o}_q$
    & \num{5.7} & \num{5.7} & \num{9.8} \\
    \bottomrule
\end{tabular}
\label{table:heat_generation_aRH}
\end{table}

\subsection{Effect of changing boundary temperatures}
The right column in Table \ref{table:consistency_all_layers} shows the entropy production in all five layers when the left boundary of the cell was kept at \qty{290}{\K}, like in the first column, while the right boundary temperature was set to \qty{291}{\K}. The changing boundary conditions did not have any impact on the entropy balance.

A negative heat flux resulted, going in the left direction, as shown in Figure S2 in the SI. The negative heat flux and the temperature gradient significantly increase the entropy production in the bulk layers, particularly in the electrodes. The entropy production in the anode and cathode surfaces is not so influenced, as the temperature jump across the interfaces is almost the same compared to the base case. A difference of one degree could easily occur by changing the electric current density, by changing external cooling facilities etc. Precisely for this reason, it appears important to know the detailed location and magnitude of the contributions to the heat flux.

\subsection{Sensitivity to surface transport properties}

The sensitivity to  surface material properties is another item of interest. In the absence of measured properties, we used a scaling factor to estimate surface properties. We examined how variations in the scaling factor influenced the temperature profile.

The surface thermal conductivities have been estimated from bulk properties using a simple scaling procedure  \cite{net_heterogen}. The procedure  involves dividing the bulk main thermal conductivity by the product of the surface thickness (\(\delta\)) and a dimensionless scaling coefficient (\(k_i\)). In this manner, the scaling coefficient becomes a factor by which the surface and bulk properties are directly compared. A scaling factor exceeding one represents an excess surface resistance, while a scaling factor below one represents a well conducting interface, not limiting to transport of heat \cite{net_heterogen}:
\begin{equation}
    \lambda_s = \frac{\lambda}{\delta k_i}
    \label{eq:correction_factor}
\end{equation}

As there are no available measurements for the surface thermal conductivities (Kapitza resistances) that we know of, we shall study the effect of a scaling factor. Spitthoff \textit{et al.} \cite{spitthoff_peltier_2021} demonstrated a substantial influence of the chosen scaling factor on both the shape of the temperature profile and the absolute temperatures. Figure S5 in the SI shows the temperature profile at the surfaces of the cell for various scaling factors, which were varied for the anode and cathode surfaces in the same manner. The blue dotted line shows the  base case scenario with $k_a= 14$ and $k_c = 110$. Additional lines were made using $k_a = k_c = 1$, 10 or 100. The scaling factor determines the deviation of the surface thermal conductivity from the corresponding value in the bulk phase. It directly influences the observed temperature jump.

It is evident that the scaling factor exerts a substantial impact on the temperature jump at the surface. As discussed in section \ref{Temperature Profile}, we can identify a correlation between the temperature jump and the temperature gradient of the adjacent bulk phase, scaled by the ratio of the thermal conductivities of the surface and the bulk phase. The finding calls for more attention to the measurement of these values.

\section{Conclusion and Perspectives}

The internal profiles of temperature, concentration and electric potential in the lithium-ion battery cell has been modelled using the theory of nonequilibrium thermodynamics (NET) as applied to heterogeneous systems \cite{net_heterogen}. We have taken Peltier heats into account, thereby extending an earlier model of the same cell \cite{spitthoff_peltier_2021}. We have shown how the cell model is thermodynamically consistent, providing a method for such checks. The NET theory gives a systematic, general approach to battery modelling. We expect that it can be useful for other electrochemical systems. Coupling effects, the trademark of the theory, are expected to be significant for porous, less well conducting electrodes with gas production or consumption. The results are sensitive to surface transport properties.

We have found that assumptions common in the lithium battery literature, to neglect or average out reversible heat effects, can lead to changes in the heat flux in the order of magnitude of \qty{30}{\%} of the value obtained from the consistent model. The thermodynamic model is essential for accurate battery temperature profile modelling.

\section*{Acknowledgement}
The authors are grateful to the Research Council of Norway, Center  of Excellence Funding Scheme, project no. 262644, PoreLab. FS is grateful to the Erasmus+ program for financially supporting the project work abroad. ØG acknowledges the Research Council of Norway for the support to the Norwegian Micro- and Nano-Fabrication Facility, NorFab, project number 295864.

\section*{Supplementary data}
Appendix: Material properties for bulk electrodes, electrolyte and electrode surfaces used in modelling.

Supplementary Information: Electric potential profile, contributions to electric potential gradient in electrolyte, simulation run and output, surface temperature jumps.

\newpage

\bibliography{references}


\clearpage

\setcounter{table}{0}
\renewcommand{\thetable}{A.\arabic{table}}

\section*{Appendix: Material Properties}\label{app:material_properties}

\begin{table}[h]
\caption{Bulk Electrode Material Properties}
\label{table:materials_electrode}
\centering
\begin{tabularx}{1.2\textwidth}{|>{\raggedright\arraybackslash}X|l|l|>{\raggedright\arraybackslash}X|l|}
    \hline
    \textbf{Property} & \textbf{Symbol} & \textbf{Unit} & \textbf{Value} & \textbf{Ref.} \\  
    \hline
    \multicolumn{5}{|c|}{Anode} \\
    \hline
    Material & \multicolumn{3}{c|}{Graphite (\ce{Li_xC6})} & \cite{spitthoff_peltier_2021, mastali_electrochemical_2016} \\
    \hline
    Thickness & $\delta$ & \unit{\um} & \num{74} & \cite{spitthoff_peltier_2021} \\
    \hline
    Thermal conductivity & $\lambda$ & \unit{\W\per\m\per\K} & \num{1.11} & \cite{thermal_conduc_temp_prof_Li_batts} \\
    \hline
    Specific Heat Capacity & $c_{p,\text{L}}$ & \unit{\J\per\mol\per\K} & \num{24.6} & \cite{cp_lithium} \\
    \hline
    Electrical conductivity & $\kappa$ & \unit{\siemens\per\m} & \num{2204} & \cite{moses_electrical_conductivities} \\
    \hline
    Peltier coefficient & $\pi^a$ & \unit{\J\per\mol} & $-1.8 \times T$ & \cite{Peltier_effects_carbon} \\
    \hline
    Diffusion coefficient & $D_{\text{L}}$ & \unit{\m\squared\per\s} & \num[print-unity-mantissa = false]{1e-11} & \cite{persson_lithium_2010} \\
    \hline
    Max. Li concentration & $c^{\text{max}}$ & \unit{\mol\per\m\cubed} & \num{30555} & \cite{mastali_electrochemical_2016} \\
    \hline
    Thermodynamic factor (SOC = 0.8) & $\Gamma$ & - & \num{0.1} & \cite{mastali_electrochemical_2016} \\
    \hline
    \multicolumn{5}{|c|}{Cathode} \\
    \hline
    Material & \multicolumn{3}{c|}{Lithium iron phosphate (\ce{LiFePO4}) LFP} & \cite{spitthoff_peltier_2021, mastali_electrochemical_2016} \\
    \hline
    Thickness & $\delta$ & \unit{\um} & \num{67} & \cite{spitthoff_peltier_2021} \\
    \hline
    Thermal conductivity & $\lambda$ & \unit{\W\per\m\per\K} & \num{0.32} & \cite{thermal_conduc_temp_prof_Li_batts} \\
    \hline
    Specific Heat Capacity & $c_{p,\text{L}}$ & \unit{\J\per\mol\per\K} & \num{24.6} & \cite{cp_lithium} \\
    \hline
    Electrical conductivity & $\kappa$ & \unit{\siemens\per\m} & \num{6.75} & \cite{moses_electrical_conductivities} \\
    \hline
    Peltier coefficient & $\pi^c$ & \unit{\J\per\mol} & $14.5 \times T$  & \cite{LFP_properties} \\
    \hline
    Diffusion coefficient & $D_{\text{L}}$ & \unit{\m\squared\per\s} & \num[print-unity-mantissa = false]{1e-11} & \cite{Morgan_2004} \\
    \hline
    Max. Li concentration & $c^{\text{max}}$ & \unit{\mol\per\m\cubed} & \num{22800} & \cite{mastali_electrochemical_2016} \\
    \hline
    Thermodynamic factor (SOC = 0.8) & $\Gamma$ & - & \num{0.5} & \cite{mastali_electrochemical_2016} \\
    \hline
\end{tabularx}
\end{table}

\begin{table}[htbp]
\caption{Electrolyte Material Properties}
\label{table:materials_electrolyte}
\centering
\begin{tabularx}{\textwidth}{|>{\raggedright\arraybackslash}X|l|l|>{\raggedright\arraybackslash}X|l|}
    \hline
    \textbf{Property} & \textbf{Symbol} & \textbf{Unit} & \textbf{Value} & \textbf{Ref.} \\  
    \hline
    Material & \multicolumn{3}{c|}{\qty{1}{M} \ce{LiPF6} in 1:1 (vol \%) EC \& DEC} & \cite{spitthoff_peltier_2021, gullbrekken_coeff} \\
    \hline
    Thickness & $\delta$ & \unit{\um} & \num{12} & \cite{spitthoff_peltier_2021} \\
    \hline
    Thermal conductivity & $\lambda$ & \unit{\W\per\m\per\K} & \num{0.2} & \cite{gullbrekken_coeff} \\
    \hline
    Electrical conductivity & $\kappa$ & \unit{\siemens\per\m} & \num{0.23} & \cite{gullbrekken_coeff} \\
    \hline
    Peltier coefficient & $\pi^e$ & \unit{\kJ\per\mol} & \num{-24.7} & \cite{gullbrekken_coeff} \\
    \hline
    \multirow{2}{*}{\begin{tabular}{@{}l@{}}Transference\\coefficients\end{tabular}} & $t_{\text{L}}$ & - & \num{-0.97} & \cite{gullbrekken_coeff} \\
    \cline{2-5}
    & $t_{\text{D}}$ & - & \num{0.9} & \cite{gullbrekken_coeff} \\
    \hline
    \multirow{2}{*}{Heats of transfer} & $q^*_{\text{L}}$ & \unit{\kJ\per\mol} & \num{1.6} & \cite{gullbrekken_coeff} \\
    \cline{2-5}
    & $q^*_{\text{D}}$ & \unit{\kJ\per\mol} & \num{0.3} & \cite{gullbrekken_coeff} \\
    \hline
    \multirow{3}{*}{\begin{tabular}{@{}l@{}}Onsager\\coefficients\end{tabular}} & $l_{\text{LL}}$ & \unit{\mol\squared\per\J\per\m\per\s} & \num{3.7e-11} $\times T$ & \cite{gullbrekken_coeff} \\
    \cline{2-5}
    & $l_{\text{DD}}$ & \unit{\mol\squared\per\J\per\m\per\s} & \num{53.7e-11} $\times T$ & \cite{gullbrekken_coeff} \\
    \cline{2-5}
    & $l_{\text{LD}}$ & \unit{\mol\squared\per\J\per\m\per\s} & \num{11.3e-11} $\times T$ & \cite{gullbrekken_coeff} \\
    \hline
    \multirow{4}{*}{\begin{tabular}{@{}l@{}}Thermodynamic \\ factor\end{tabular}} & $\Gamma_{\text{LL}}$ & - & \num{1.45} & \cite{gullbrekken_coeff} \\
    \cline{2-5}
    & $\Gamma_{\text{DD}}$ & - & \num{-0.29} & \cite{gullbrekken_coeff} \\
    \cline{2-5}
    & $\Gamma_{\text{LD}}$ & - & \num{-0.98} & \cite{gullbrekken_coeff} \\
    \cline{2-5}
    & $\Gamma_{\text{DL}}$ & - & \num{1.23} & \cite{gullbrekken_coeff} \\
    \hline
\end{tabularx}
\end{table}

\begin{table}[htb]
\caption{Electrode Surface Material Properties}
\label{table:materials_surface}
\centering
\begin{threeparttable}
\begin{tabularx}{1.2\textwidth}{|>{\raggedright\arraybackslash}X|l|l|>{\raggedright\arraybackslash}X|l|}
    \hline
    \textbf{Property} & \textbf{Symbol} & \textbf{Unit} & \textbf{Value} & \textbf{Ref.} \\  
    \hline
    \multicolumn{5}{|c|}{Anode Surface} \\
    \hline
    Material & \multicolumn{3}{c|}{\ce{Li2CO3}, \ce{LiF}, graphite} & \cite{spitthoff_peltier_2021} \\
    \hline
    Thickness & $\delta$ & \unit{\nm} & \num{50} & \cite{Wang_review_modeling_SEI} \\
    \hline
    Thermal conductivity (bulk) & $\lambda$ & \unit{\W\per\m\per\K} & \num{0.65} (\ce{Na2CO3})\tnote{1} & \cite{molec_sim_thermal_transp_alka_carb} \\
    \hline
    Surface thermal conductivity  & $\lambda^s$ & \unit{\W\per\m\squared\per\K} & \num{928571}\tnote{2} & - \\
    \hline
    Open circuit potential (SOC = 0.8) & $U_{\text{eq}}$ & \unit{\V} (vs. \ce{Li}/\ce{Li+}) & \num{0.10} & \cite{Allart_2018} \\
    \hline
    Exchange current density & $j_{0}$ & \unit{\A\per\m\squared} & \num{0.87} & \cite{Ge_2017} \\
    \hline
    Peltier heat &  $\Pi^{s,a}$ & \unit{\kJ\per\mol} & \unit{-104} & \cite{gunnarshaug_reviewreversible_2021} \\
    \hline
    Scaling factor & $k$ & - & \num{14} & \cite{spitthoff_peltier_2021} \\
    \hline
    \multicolumn{5}{|c|}{Cathode Surface} \\
    \hline
    Material & \multicolumn{3}{c|}{Carbon coating} & \cite{opt_carbon_coating_LFP} \\
    \hline
    Thickness & $\delta$ & \unit{\nm} & \num{10} & \cite{spitthoff_peltier_2021} \\
    \hline
    Thermal conductivity (bulk) & $\lambda$ & \unit{\W\per\m\per\K} & \num{1.11} & \cite{thermal_conduc_temp_prof_Li_batts} \\
    \hline
    Surface thermal conductivity  & $\lambda^s$ & \unit{\W\per\m\squared\per\K} & \num{1009091}\tnote{1} & - \\
    \hline
    Open circuit potential (SOC = 0.8) & $U_{\text{eq}}$ & \unit{\V} (vs. \ce{Li}/\ce{Li+}) & \num{3.45} & \cite{OCV_LFP} \\
    \hline
    Exchange current density & $j_{0}$ & \unit{\A\per\m\squared} & \num{1.7} & \cite{charg_transf_kinet_LFP} \\
    \hline
    Peltier heat & $\Pi^{s,c}$ & \unit{\kJ\per\mol} & \num{122} & \cite{gunnarshaug_reviewreversible_2021} \\
    \hline
    Scaling factor & $k_i$ & - & \num{110} & \cite{spitthoff_peltier_2021} \\
    \hline
\end{tabularx}
\begin{tablenotes}
    \item[1] Assumed that thermal conductivity of \ce{Na2CO3} is similar to \ce{Li2CO3}.
    \item[2] Calculated from Eq. \eqref{eq:correction_factor}.
\end{tablenotes}
\end{threeparttable}
\end{table}

\newpage

\end{document}